\documentclass[prb,aps,superscriptaddress,reprint]{revtex4-2}

\usepackage{graphicx}
\usepackage{amsmath}
\usepackage[english]{babel}
\usepackage{units}
\usepackage[colorlinks=true,urlcolor=blue,linkcolor=blue,citecolor=blue]{hyperref}
\usepackage{mathptmx}
\usepackage{xcolor}

\usepackage[noindentafter]{titlesec}
\titleformat{\section}{\raggedright\bfseries}{}{0pt}{}
\titlespacing{\section}{0pt}{10pt}{0pt}
\titleformat{\subsection}{\raggedright\bfseries}{}{0pt}{}
\titlespacing{\subsection}{0pt}{7pt}{0pt}

\begin{document}
\title{Twist angle dependent interlayer transfer of valley polarization from excitons to free charge carriers in WSe$_2$/MoSe$_2$ heterobilayers}

\author{Frank Volmer}
\affiliation{2nd Institute of Physics and JARA-FIT, RWTH Aachen University, 52074 Aachen, Germany}
\affiliation{AMO GmbH, Advanced Microelectronic Center Aachen (AMICA), 52074 Aachen, Germany}

\author{Manfred Ersfeld}
\affiliation{2nd Institute of Physics and JARA-FIT, RWTH Aachen University, 52074 Aachen, Germany}

\author{Paulo E. Faria~Junior}
\affiliation{Institut für Theoretische Physik, Universität Regensburg, D-93040 Regensburg, Germany}

\author{Lutz Waldecker}
\affiliation{2nd Institute of Physics and JARA-FIT, RWTH Aachen University, 52074 Aachen, Germany}
\affiliation{Department of Applied Physics, Stanford University, 348 Via Pueblo Mall, Stanford, CA 94305, USA}

\author{Bharti Parashar}
\affiliation{Peter Gr\"unberg Institute (PGI-6), Forschungszentrum J\"ulich GmbH, 52428 J\"ulich, Germany}

\author{Lars Rathmann}
\affiliation{2nd Institute of Physics and JARA-FIT, RWTH Aachen University, 52074 Aachen, Germany}

\author{Sudipta Dubey}
\affiliation{2nd Institute of Physics and JARA-FIT, RWTH Aachen University, 52074 Aachen, Germany}

\author{Iulia Cojocariu}
\affiliation{Peter Gr\"unberg Institute (PGI-6), Forschungszentrum J\"ulich GmbH, 52428 J\"ulich, Germany}

\author{Vitaliy Feyer}
\affiliation{Peter Gr\"unberg Institute (PGI-6), Forschungszentrum J\"ulich GmbH, 52428 J\"ulich, Germany}
\affiliation{Fakult\"at f\"ur Physik and Center for Nanointegration Duisburg-Essen (CENIDE), Universität Duisburg-Essen, D-47048 Duisburg, Germany}

\author{Kenji Watanabe}
\affiliation{Research Center for Functional Materials, National Institute for Materials Science, Tsukuba 305-0044, Japan}

\author{Takashi Taniguchi}
\affiliation{International Center for Materials Nanoarchitectonics, National Institute for Materials Science, Tsukuba 305-0044, Japan}

\author{Claus M. Schneider}
\affiliation{Peter Gr\"unberg Institute (PGI-6), Forschungszentrum J\"ulich GmbH, 52428 J\"ulich, Germany}
\affiliation{Fakult\"at f\"ur Physik and Center for Nanointegration Duisburg-Essen (CENIDE), Universität Duisburg-Essen, D-47048 Duisburg, Germany}

\author{Lukasz Plucinski}
\affiliation{Peter Gr\"unberg Institute (PGI-6), Forschungszentrum J\"ulich GmbH, 52428 J\"ulich, Germany}

\author{Christoph Stampfer}
\affiliation{2nd Institute of Physics and JARA-FIT, RWTH Aachen University, 52074 Aachen, Germany}
\affiliation{Peter Gr\"unberg Institute (PGI-9), Forschungszentrum J\"ulich, 52425 J\"ulich, Germany}

\author{Jaroslav Fabian}
\affiliation{Institut für Theoretische Physik, Universität Regensburg, D-93040 Regensburg, Germany}

\author{Bernd Beschoten}
\email{bernd.beschoten@physik.rwth-aachen.de}
\affiliation{2nd Institute of Physics and JARA-FIT, RWTH Aachen University, 52074 Aachen, Germany}
\affiliation{JARA-FIT Institute for Quantum Information, Forschungszentrum J\"ulich GmbH and RWTH Aachen University, 52074 Aachen, Germany}

\begin{abstract}
Transition metal dichalcogenides (TMDs) have attracted much attention in the fields of valley- and spintronics due to their unique property of forming valley-polarized excitons when illuminated by circularly polarized light. In TMD-heterostructures it was shown that these electron-hole pairs can scatter into valley-polarized interlayer exciton states, which exhibit long lifetimes and a unique twist-angle dependence. However, the question how to create a valley polarization of free charge carriers in these heterostructures after a valley selective optical excitation is unexplored, despite its relevance for opto-electronic devices. Here, we identify an interlayer transfer mechanism in twisted WSe$_2$/MoSe$_2$ heterobilayers that transfers the valley polarization from excitons in WSe$_2$ to free charge carriers in MoSe$_2$ with valley lifetimes of up to \unit[12]{ns}. This mechanism is most efficient at large twist angles, whereas the valley lifetimes of free charge carriers are surprisingly short for small twist angles, despite the occurrence of interlayer excitons. 
\end{abstract}

\maketitle
\section{Introduction}
Using excitons in transition metal dichalcogenides (TMDs) to store and manipulate information in their valley degree of freedom is argued to be an appealing alternative to charge-based information processing~\cite{NatureReviewsMaterials.7.449,NatureReviewsMaterials.1.16055,npj2DMaterialsandApplications.2.29,LightScienceApplications.1.72,NanoResearch.12.2695}. Especially heterostructures made from two different, semiconducting TMDs, like WSe$_2$/MoSe$_2$-heterobilayers, are of great interest as excitons in these heterobilayers can be controlled by electrical means~\cite{Science.366.870,NaturePhotonics.13.131,PRB.101.121404,NatureComm.11.2640} and exhibit recombination and valley lifetimes in the nanosecond range~\cite{Science.351.688,NatComm.6.6242,NanoLetters.17.5229,NatureComm.9.753,NatureCommunications.8.1551}. The possibility to manipulate these excitons via the twist angle between the two TMD layers~\cite{ACSNano.15.14725,NatureMaterials.19.617,PhysRevLett.126.047401,ACSNano.11.4041} and the appearance of unique spin-valley and many-body physics~\cite{PhysRevResearch.2.042044,NatureMaterials.19.624,NaturePhysics.15.1140,Nature.580.472,npj2DMaterialsandApplications.4.8,PRB.101.235408} made TMD-based heterobilayers even more interesting in the emerging fields of both valleytronics and twistronics~\cite{NatureReviewsMaterials.7.449,NatureReviewsMaterials.1.16055,npj2DMaterialsandApplications.2.29}. However, most studies focus on pure exciton physics and do not address the questions, if and how valley-polarized excitons in such heterobilayers can transfer their valley polarization to free charge carriers in either their conduction or valence bands, and if such a transfer mechanism can be controlled by parameters like the twist-angle or the position of the Fermi level. Answers to these questions are crucial for the realization of opto-valleytronic devices, in which the optically excited valley polarization may be extracted and measured by electrical means.

In this article, we report on an optical excitation mechanism that transfers the valley polarization from excitons in WSe$_2$ to free charge carriers in MoSe$_2$ via an interlayer charge transfer in twisted WSe$_2$/MoSe$_2$-heterobilayers. We show that this transfer of valley polarization significantly depends on both the twist angle and the position of the Fermi level. In devices with large twist angles, the interlayer valley transfer becomes most efficient when crossing the band edges of both the MoSe$_2$ valence and conduction band. This behavior is explained by twist angle-dependent scattering mechanisms that involve the Q- and $\Gamma$-valleys, where the latter is probed by angle-resolved photoemission spectroscopy (ARPES). To measure both the valley polarization and the respective valley lifetimes, we employ an all-optical, time-resolved Kerr rotation (TRKR) measurement technique that has proven to be a powerful tool to investigate the valley dynamics~\cite{NanoLetters.21.7123,2DMaterials.8.025011,NanoLetters.19.4083,NanoLett.16.5010,NatPhys.11.830,ReviewScientificInstruments.92.113904,2DMaterials.5.011010,PhysRevB.90.161302,NatureComm.7.12715,PhysRevB.95.235408,NatureCommunications.13.4997}. In TRKR, circularly polarized laser pump pulses are typically used to resonantly excite valley-polarized excitons, while their temporal dynamics is being probed by linearly polarized laser probe pulses measuring the Kerr rotation angle. If the energy of the probe pulse is tuned to the trion (charged exciton) energy of the TMD material, TRKR is able to detect a valley polarization of free charge carriers in the corresponding TMD layer, as demonstrated in Ref.~\cite{NanoLetters.20.3147} and discussed in the Supplementary Information~\cite{Supplement}. Therefore, for all measurements in this study, the pump and probe energies are the trion energies of the specified TMD unless stated otherwise.

\begin{figure*}[tb]
	\includegraphics[width=\linewidth]{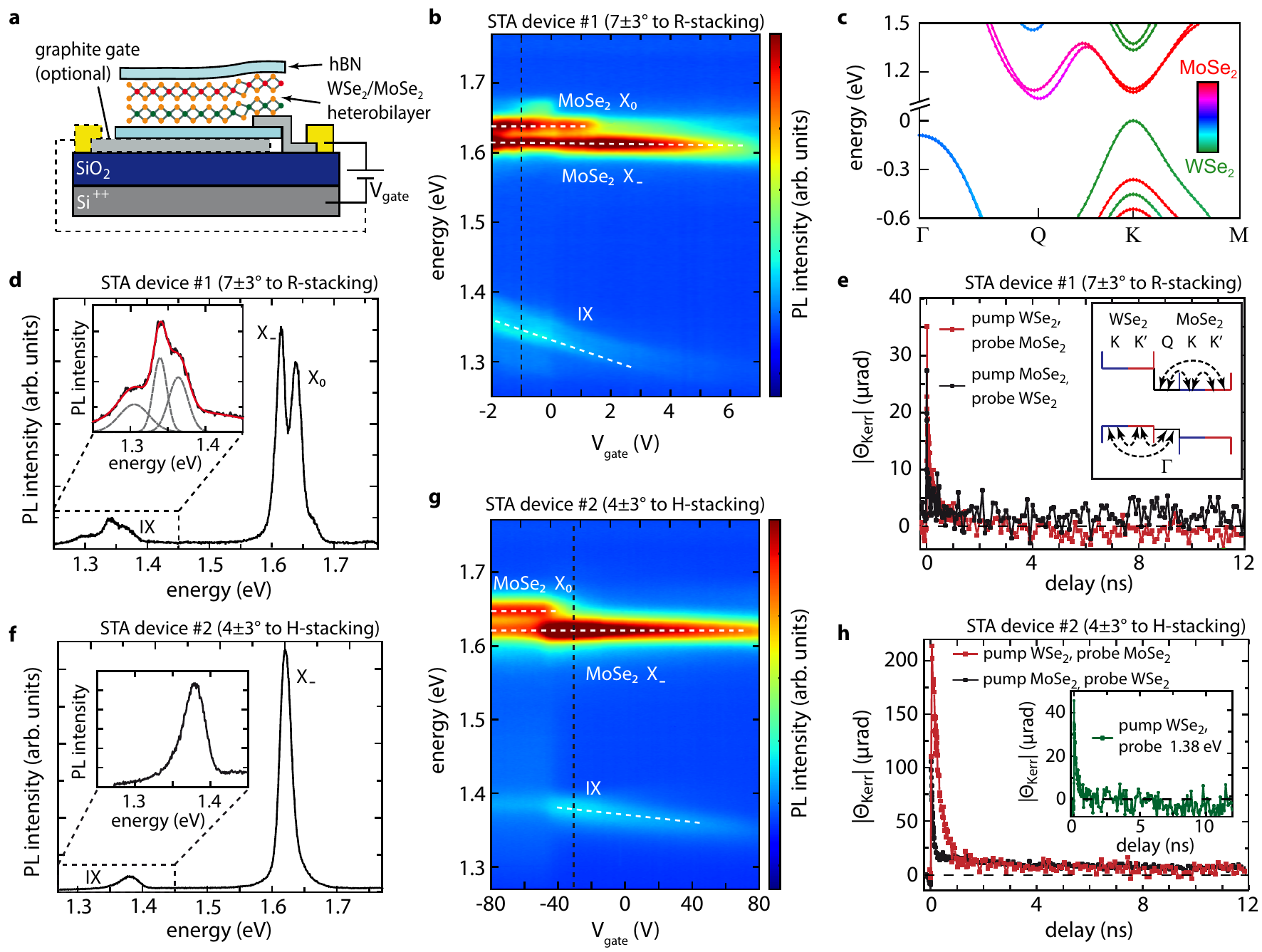}
	\caption{\textbf{Photoluminescence and time-resolved Kerr rotation measurements on WSe$_2$/MoSe$_2$ heterobilayers with small twist angles towards a crystallographic stacking order.} \textbf{a} Schematic device structure of the heterobilayer devices. Only STA device~\#1 and LTA devices~\#1 and \#3 have a graphite gate, whereas STA device~\#2 is gated via the Si$^{++}$/SiO$_2$ (\unit[285]{nm}) wafer. \textbf{c} Calculated band structure in case of the RhX-stacking (i.e.\ a twist angle of zero), in which the amount of hybridization is color-coded. The bands at the K-point only show negligible hybridization and a type-II band alignment. The bands at the Q- and $\Gamma$-valleys are strongly hybridized and, depending on parameters like stacking order or twist angle, shift in energy with respect to the band maxima at the K-point (see Supplementary Information~\cite{Supplement}). \textbf{b,g} Photoluminescence (PL) spectra as a function of gate voltage $V_\text{gate}$ for two different devices plotted with a logarithmic color scale and \textbf{d,f} respective line-cuts plotted on a linear scale along the black dashed lines in \textbf{b} and \textbf{g} (white dashed lines are guides to the eye). The features with the highest PL intensities are the intralayer neutral exciton (X$_0$) and the trion (X$_-$) emission of MoSe$_2$. For STA device \#1, the interlayer exciton (IX) emission consists of several sub-peaks (see fits in the inset of \textbf{d}), which can be attributed to transitions involving Q- and $\Gamma$-valleys. As these features are absent in STA device \#2 (see inset in \textbf{f}), we assume that in STA device \#2 both Q- and $\Gamma$-valleys are energetically further away from the K-valleys decreasing phonon-assisted scattering between the valleys (see schematic in the inset of \textbf{e}). This assumption is consistent to the TRKR data shown in \textbf{e,h}. For STA device \#1, we observe surprisingly short lifetimes of around \unit[200]{ps} and extremely small amplitudes. Instead, Kerr rotation lifetimes and amplitudes of  STA device \#2 are one order of magnitude larger. All measurements were conducted at \unit[10]{K}, pump and probe energies are the trion energies of the specified TMD unless stated otherwise.}
	\label{fig1}
\end{figure*}

We investigate five WSe$_2$/MoSe$_2$-heterobilayer devices (see schematic device layout in Fig.\,\ref{fig1}a and details on the device fabrication in the Method section) with varying twist angles, which were determined by polarization-dependent measurements of the second harmonic generation (see Supplementary Information~\cite{Supplement} and~\cite{ACSNano.8.2951}). The devices can be divided into two subgroups, which, as we will show, have substantially different valley dynamics: The first subgroup has a small twist angle (STA) towards a well-defined crystallographic stacking order (an R-type stacking order for a twist angle of zero, an H-type stacking order for a twist angle of $\unit[60]{^\circ}$~\cite{ACSNano.14.4550,PRB.101.235408,ACSNano.11.4041,Nanomaterials.13.1187}), while the second subgroup has a large twist angle (LTA) with respect to these stacking orders.  

\section{Results}
\subsection{Heterobilayers with small twist angles}
We start our discussion with a WSe$_2$/MoSe$_2$ heterobilayer that has a small twist angle of $\unit[7\pm3]{^\circ}$ towards an R-type stacking order (STA device \#1). A photoluminesence (PL) measurement as a function of the gate voltage $V_\text{gate}$ is shown in Fig.\,\ref{fig1}b, whereas Fig.\,\ref{fig1}d depicts a line-cut along the black dashed line of Fig.\,\ref{fig1}b. The two features with the highest PL intensities are the intralayer neutral exciton (X$_0$) and trion (X$_-$) emission of MoSe$_2$. One observation that will get relevant as soon as we compare this device to ones that have larger twist angles is the absence of any intralayer PL emission from WSe$_2$. We note that even the PL emission from MoSe$_2$ is significantly reduced compared to the monolayer case, which can be attributed to an ultrafast charge transfer from intralayer into interlayer excitons~\cite{NatNanotechnol.13.994,ACSNano.15.14725,NatureNanotechnology.9.682,ACSNano.11.12020}. In fact, a majority of the photo-excited electron-hole pairs must have been scattered into the interlayer exciton state (IX) to explain the appearance of its emission in Fig.\,\ref{fig1}b despite its very small oscillator strength \cite{CommunicationsPhysics.2.23,PRB.97.165306}. The interlayer exciton appears between \unit[1.3]{eV} and \unit[1.4]{eV} and shows the expected gate-dependent energy shift due to its dipole moment \cite{NaturePhotonics.13.131,NatureMaterials.19.624} (dashed white lines in Fig.\,\ref{fig1}b are guides to the eye).

In this device, the interlayer exciton emission consists of several sub-peaks (see fits in the inset of Fig.\,\ref{fig1}d) which exact origins are still a debated topic \cite{J.Phys.Condens.Matter.32.333002}. Some studies argue that these sub-features can be related to transitions involving Q- and $\Gamma$-valleys \cite{NanoLetters.17.5229,ACSNano.11.4041,PRB.97.165306,ACSNano.12.2498,ACSNano.12.4719,PhysRevResearch.2.042044}. If this is the case, the appearance of these sub-features would imply a strong hybridization between the two TMD layers. To address this, we refer to band structure calculations that predict that the bands at the K-point of the heterobilayer only show negligible signs of hybridization, i.e.\ that the conduction band minimum at the K-point of the heterobilayer mainly consists of MoSe$_2$ states, whereas the valence band maximum at the K-point mainly consists of WSe$_2$ states, leading to a type-II band alignment~\cite{PRB.97.165306,Phys.Rev.B.97.205417,ACSNano.12.2498,ACSNano.12.4719} (see Fig.\,\ref{fig1}c for the calculated band structure in case of the RhX-stacking). On the other hand, the bands at both the Q- and $\Gamma$-valleys are strongly hybridized and, depending on parameters like stacking order and twist angle, may shift in energy towards the band maxima at the K-point (see Supplementary Information~\cite{Supplement} for more calculations). Interestingly, different studies significantly differ in the prediction of these energy shifts. Some of them claiming a transition to an indirect semiconductor as either the Q-valley shifts below the K-point's conduction band minimum, or the $\Gamma$-valley shifts above the K-point's valence band maximum \cite{PRB.97.165306,Phys.Rev.B.97.205417,ACSNano.12.2498,ACSNano.12.4719,PhysRevResearch.3.043217}. As we observe the sub-features in the interlayer exciton emission, we assume that both the Q- and $\Gamma$-valleys are energetically close enough to the K-valleys that phonon-assisted scattering can take place between these valleys (see schematic in the inset of Fig.\,\ref{fig1}e).
 
This finding is important for the interpretation of the time-resolved Kerr rotation data shown in Fig.\,\ref{fig1}e, where we plot the Kerr rotation amplitude $\Theta_\text{Kerr}$ vs the time delay between pump (circularly polarized) and probe (linearly polarized) pulses. Regardless where we set the pump and probe energies or which gate voltage we apply, we observe surprisingly short lifetimes of around \unit[200]{ps} and small amplitudes of $\Theta_\text{Kerr} < \unit[40]{\mu rad}$. The only Kerr rotation signal with a good enough signal-to-noise ratio to be analyzed was obtained by tuning the pump and probe energies to the trion energies of WSe$_2$ and MoSe$_2$ (see~Fig.\,\ref{fig1}e). Instead, no signal was detected at the interlayer exciton energies (not shown). This is highly surprising in two ways: On the one hand, these values fall significantly short compared to lifetimes of up to nanoseconds and amplitudes of up to several hundreds of $\mu$rad that are typically observed in TRKR measurements on monolayer TMDs~\cite{NanoLetters.21.7123,2DMaterials.8.025011,NanoLetters.19.4083,NanoLett.16.5010,NatPhys.11.830,ReviewScientificInstruments.92.113904,2DMaterials.5.011010,PhysRevB.90.161302,NatureComm.7.12715,PhysRevB.95.235408,NanoLetters.20.3147}. On the other hand, and even more surprising, is the fact that the lowest valley lifetimes of interlayer excitons measured by time- and helicity-resolved PL measurements in WSe$_2$/MoSe$_2$-heterobilayers are also in the nanosecond range~\cite{Science.351.688,NatComm.6.6242,NanoLetters.17.5229,NatureComm.9.753,NatureCommunications.8.1551}. Although we observe interlayer excitons in PL (see Fig.\,\ref{fig1}b), we conclude that TRKR is in all likelihood unable to directly detect the interlayer exciton valley polarization (see further discussion in the Supplemental Material~\cite{Supplement}). Moreover, the short lifetimes in TRKR probed at the trion energies (Fig.\,\ref{fig1}e) imply that any valley polarization of free charge carriers relaxes quite fast in this heterobilayer. We attribute this to the aforementioned presence of the $\Gamma$- and Q-valleys, which provide additional scattering channels for a valley polarization as soon as these valleys are energetically in the phonon-assisted scattering range (see inset in Fig.\,\ref{fig1}e). This indicates the importance of these valleys in the understanding of the overall spin and valley dynamics of free charge carriers in heterobilayers that have small twist angles near to a crystallographic stacking order, which is in accordance with previous theoretical and experimental studies based on optical measurement techniques other than TRKR~\cite{NatNanotechnol.13.994,LightScienceApplications.1.72}.

\begin{figure*}[tb]
	\includegraphics[width=\linewidth]{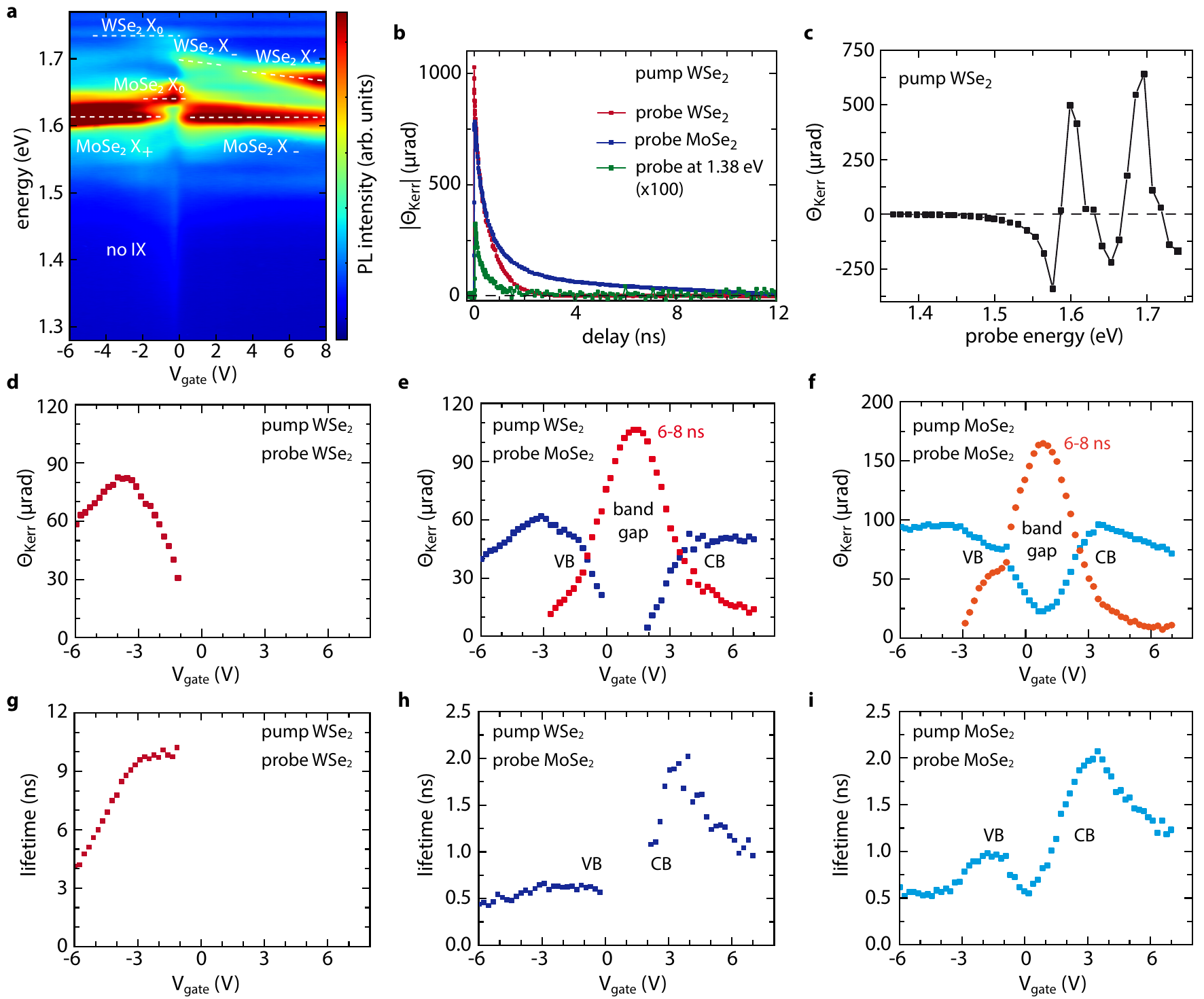}
	\caption{\textbf{Photoluminescence and time-resolved Kerr rotation measurements on a WSe$_2$/MoSe$_2$ heterobilayer with a large twist angle of $\unit[23\pm3]{^\circ}$ (LTA device~\#1)} \textbf{a} The gate-dependent PL spectra that are plotted on a logarithmic color scale show no interlayer exciton emission, but instead the intralayer exciton ($X_0$) and trion peaks ($X_-$ and $X'_-$) of WSe$_2$. The latter were absent in Figs.\,\ref{fig1}\textbf{b} and \ref{fig1}\textbf{g} because of hybridization-induced scattering channels in devices with small twist angles. \textbf{b} Conversely, the appearance of the WSe$_2$ intralayer emission in LTA device \#1 implies the suppression of these scattering channels, resulting in TRKR lifetimes and amplitudes that are much larger than the ones of the previous devices. \textbf{c} An energy scan, where the pump pulse energy is set to the trion energy of WSe$_2$ and the probe pulse energy is varied, only shows resonances at the bright intralayer exciton and trion energies. A Kerr rotation signal at the interlayer exciton energy, which is scaled-up by a factor of 100 in \textbf{b} for better visibility, is only due to the long tail of the bright exciton resonances. \textbf{d-f} Gate-dependent TRKR amplitudes and \textbf{g-i} lifetimes for different combinations of pump and probe energies, which reveal two different types of signals. The first signal (reddish colors) appears if the Fermi level lies in the band gap region of the probed TMD layer and can be linked to defect-bound exciton states. The second signal (blueish colors) appears as soon as the Fermi level is tuned either into the conduction (CB) or valence band (VB) of MoSe$_2$ (compare to the appearance of the trion features in \textbf{a}) and can be attributed to a valley polarization of free charge carriers (see explanation in the text). All measurements were conducted at \unit[10]{K}, pump and probe energies are the trion energies of the specified TMD unless stated otherwise.}
	\label{fig2}
\end{figure*}

To support this notion, we studied another WSe$_2$/MoSe$_2$ heterobilayer (STA device \#2) with a twist angle of $\unit[56\pm3]{^\circ}$, which is a small twist angle of $\unit[4\pm3]{^\circ}$ to an H-type stacking order. This device also shows interlayer excitons in the gate-dependent PL map (see Fig.\,\ref{fig1}g), but without any well separated sub-peaks (see inset in Fig.\,\ref{fig1}f, which is a line-cut along the black dashed line in Fig.\,\ref{fig1}g). This may imply that the Q- and $\Gamma$-valleys are energetically further separated from the K-valleys and therefore no longer play a key role in the interlayer exciton emission. The less pronounced hybridization compared to STA device \#1 might be due to the different stacking order (see also our calculations in the Supplementary Information~\cite{Supplement}), an increased separation between the layers, or strain~\cite{Zollner2019PRB,Kunstmann2018NatPhys}. If we assume a larger energy separation between the K-valleys and the Q- or $\Gamma$-valleys, we expect a less pronounced scattering rate between these valleys. This is in complete accordance to the results of the TRKR measurements on this device, which yield Kerr rotation lifetimes and amplitudes which are one order of magnitude larger than the ones of STA device \#1 (compare Figs.\,\ref{fig1}e and \ref{fig1}h). Setting the energies of the pump and probe laser pulses to the trion energies of the respective TMD layers, we now observe a fast decaying Kerr rotation signal of several hundreds of picoseconds and a second smaller signal with a decay in the nanosecond range. Interestingly, we even observe a small Kerr rotation signal when the probe pulse is set to the interlayer exciton energy (see the inset of Fig.\,\ref{fig1}h). However, it must be the topic of further studies to clarify, if this signal really originates from the interlayer exciton or is rather caused by a low-energy tail of a bright, intralayer exciton resonance (see e.g.\ the discussions to Figs.\,\ref{fig2}b and \ref{fig2}c later on).

\subsection{Heterobilayers with large twist angles}
An even smaller impact of the Q- and $\Gamma$-valleys due to a larger energy separation to the K-valleys is expected for twist angles that are far away from a crystallographic stacking order (see calculations in the Supplementary Information~\cite{Supplement}). Therefore, we now focus on a heterobilayer with a large twist angle of $\unit[23\pm3]{^\circ}$ (LTA device \#1). The gate-dependent PL map of this device is shown in Fig.\,\ref{fig2}a and reveals a significantly different coupling between the two TMD layers. This is evident not only in the absence of an interlayer exciton emission~\cite{ACSNano.11.4041}, but also in the appearance of the intralayer exciton ($X_0$) and trion peaks ($X_-$ and $X'_-$) of WSe$_2$~\cite{RevModPhys.90.021001,NatureCommunications.9.3718,OpticalMaterialsExpress.10.1273}, which were completely quenched in the previous two devices (compare Fig.\,\ref{fig2}a to Figs.\,\ref{fig1}b and \ref{fig1}g). It is important to note that many studies about WSe$_2$/MoSe$_2$-heterobilayers that are showing interlayer exciton emission show a much stronger quenching of WSe$_2$ intralayer excitons compared to the quenching of MoSe$_2$ intralayer excitons~\cite{ACSNano.12.4719,npj2DMaterialsandApplications.4.8,NatureComm.9.753,NatComm.6.6242,NaturePhotonics.13.131,PhysRevLett.126.047401}. This implies the existence of fast, hybridization-induced scattering channels for photo-excited electron-hole pairs away from the K-valleys of WSe$_2$ in heterobilayers with small twist angles (see Supplemental Material for a more detailed discussion~\cite{Supplement}). Conversely, the appearance of the WSe$_2$ intralayer emission in LTA device \#1 implies the suppression of these scattering channels for larger twist angles and therefore a significant decrease in the coupling strength between the TMD layers.

The conclusion of a decreased coupling is further supported by the TRKR measurements on LTA device \#1 (Fig.\,\ref{fig2}b) that exhibit lifetimes and amplitudes that are much higher than those of the STA devices (compare Fig.\,\ref{fig2}b to Figs.\,\ref{fig1}e and \ref{fig1}h). In fact, amplitudes of hundreds of $\mu$rad and lifetimes in the nanosecond range are reminiscent to values measured on monolayer TMDs~\cite{NanoLetters.21.7123,2DMaterials.8.025011,NanoLetters.19.4083,NanoLett.16.5010,NatPhys.11.830,ReviewScientificInstruments.92.113904,2DMaterials.5.011010,PhysRevB.90.161302,NatureComm.7.12715,PhysRevB.95.235408,NanoLetters.20.3147}. The energy scan in Fig.\,\ref{fig2}c, in which the pump pulse energy was fixed at the trion energy of WSe$_2$ and the probe pulse was tuned over the whole energy range of the PL measurements, only shows resonances at the bright intralayer exciton and trion energies. A small Kerr rotation signal at the interlayer exciton energy (between \unit[1.3]{eV} and \unit[1.4]{eV} in case of the STA devices), which is scaled-up by a factor of 100 in Fig.\,\ref{fig2}b for better visibility, is only due to the long tail of the bright exciton resonances in Fig.\,\ref{fig2}c.

Interestingly, two distinctly different TRKR signals can be revealed in the heterobilayer by tuning the Fermi level via a gate voltage and by varying the energies of both pump and probe pulses (see Figs.\,\ref{fig2}d to \ref{fig2}f for the respective amplitudes $\Theta_{\text{Kerr}}$ of the TRKR signals and the Supporting Information of Ref.~\cite{NanoLetters.20.3147} for the used fitting procedure). The occurrence of these two signals depends on whether the Fermi level is tuned either in the band gap of the probed TMD layer (reddish colors in Figs.\,\ref{fig2}d to \ref{fig2}f) or in the valence or conduction band (bluish colors). For example, if we set the energies of both the pump and probe pulses to the trion energy of WSe$_2$ (Fig.\,\ref{fig2}d), a Kerr rotation signal with a lifetime between \unit[4]{ns} and \unit[10]{ns} (Fig.\,\ref{fig2}g) appears as soon as the trion emission of WSe$_2$ in Fig.\,\ref{fig2}a vanishes for gate voltages smaller than \unit[0]{V}, i.e.\ as soon as we leave the conduction band and enter the band gap of WSe$_2$. If the probe pulse energy is instead tuned to the trion energy of MoSe$_2$ (see Figs.\,\ref{fig2}e and \ref{fig2}f, where we used different pump energies), a TRKR signal with similarly long lifetimes appears in the gate voltage range that also shows the appearance of the neutral exciton ($X_0$) emission of MoSe$_2$ in Fig.\,\ref{fig2}a, which in turn marks the gate range of the band gap in MoSe$_2$. Similar long-lived states within the gate voltage range, in which emission of neutral excitons predominates, have been reported previously in both WSe$_2$/MoSe$_2$ heterobilayers \cite{Science.360.893} and WSe$_2$ monolayer devices~\cite{NanoLetters.20.3147,PhysRevMaterials.5.044001}.

The corresponding amplitude of this long-lived signal is maximal within the band gap region and falls as soon as the Fermi level enters either the conduction band (VB) or valence band (VB) of MoSe$_2$, where the band edges are determined by the appearance of the negatively or positively charged trion emission in Fig.\,\ref{fig2}a, respectively. Small shifts in the gate-dependent positions of the features in Fig.\,\ref{fig2}a compared to Figs.\,\ref{fig2}d to \ref{fig2}f are expected due to photo-induced gate screening effects that depend on both the energy and intensity of the used laser systems, which are different for TRKR and PL measurements (see Ref.~\cite{PhysStatusSolidiRRL.14.2000298} and the Supporting Information of Ref.~\cite{NanoLetters.20.3147}). As we discuss further below in more detail, the band gap related TRKR signal results from defect-bound exciton states within the band gap. Accordingly, the corresponding TRKR signal has a lifetime similar to the recombination time of these bound exciton states, which spans the nanosecond to microsecond range at cryogenic temperatures~\cite{PhysRevB.96.121404,PhysRevLett.121.057403,NatureCommunications.12.871}.

A distinctly different TRKR signal in the WSe$_2/$MoSe$_2$ heterobilayer appears when the Fermi level gets tuned into either the valence or the conduction band of MoSe$_2$ and when the probe energy is tuned to MoSe$_2$ (see blueish colors in Figs.\,\ref{fig2}e and \ref{fig2}f). The initial increase of the Kerr rotation amplitude indicates that this signal is linked to a valley-polarization of free charge carriers. Interestingly, such a gate-dependent TRKR signal is completely missing if both the pump and probe energies are set to the WSe$_2$ energy (Fig.\,\ref{fig2}d)~\footnote{For positive gate voltages, for which the trion emission of WSe$_2$ is present in Fig.\,\ref{fig2}a, we only observe a short-lived signal in the lower picosecond range that we assign to bright exciton recombinations. As bright excitons are not the scope of our study, we do not include the corresponding signal in Figs.\,\ref{fig2}d to \ref{fig2}f.}. Remarkably, this result is exactly opposite to what was found in the respective monolayer cases, as some of us demonstrated in Ref.~\cite{NanoLetters.20.3147}. There, a valley polarization of free charge carriers could only be achieved in WSe$_2$, where the polarization mechanism was explained by an intervalley scattering channel via dark trion states, which are not available in MoSe$_2$ monolayers~\cite{RevModPhys.90.021001}. Comparing Figs.\,\ref{fig2}d and \ref{fig2}e instead reveals the existence of a mechanism that transfers the valley polarization from excitons in WSe$_2$ to free charge carriers in MoSe$_2$ via an interlayer charge transfer.

\begin{figure*}[tb]
	\includegraphics[width=\linewidth]{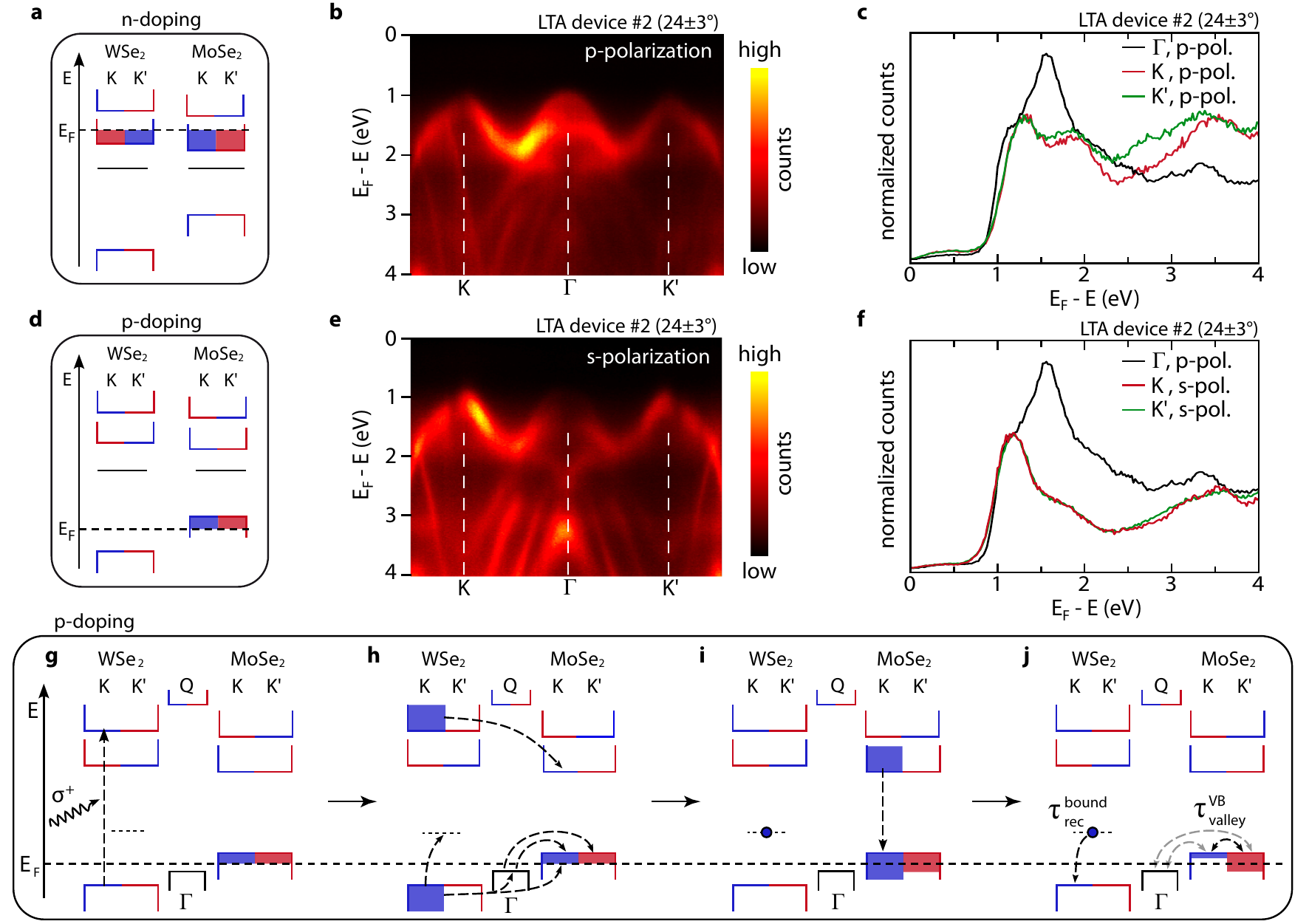}
	\caption{\textbf{Band structure at the K-, Q- and $\Gamma$-valleys for LTA devices and the resulting transfer mechanism of a valley polarization from excitons to free charge carriers.} \textbf{a,d} Band structure of LTA device \#1 at the K-points derived from the gate-dependence of the bright exciton emission in Fig.\,\ref{fig2}\textbf{a}. Blue and red colors represent spin-up and spin-down states, respectively. Sightly opaque rectangles represent the filling of the bands with charge carriers. \textbf{a} The fact that the negatively charged trion features of WSe$_2$ and MoSe$_2$ appear almost at the same gate voltage in Fig.\,\ref{fig2}\textbf{a} hints to an alignment of the respective conduction band minima to each other. This deviation from the expected type-II band alignment may be caused by a combination of the large twist angle together with a slight $n$-doping of both TMD layers (black solid lines below the conduction bands in \textbf{a,d} represent donor states). \textbf{d} For negative gate voltages, the smaller band gap of MoSe$_2$ prevents the Fermi level to enter the valence band of WSe$_2$ within the applicable gate voltage range, explaining the absence of the positively charged trion emission of WSe$_2$. \textbf{b,e} ARPES measurements of LTA device \#2 for both s- and p-polarized light and \textbf{c,f} linecuts that are normalized and averaged over a small wave-vector range at the K-, K'-, and $\Gamma$-points demonstrate that in this device the $\Gamma$-valley is energetically close to the K-valleys. \textbf{g-j} Model that is based on the derived and measured band structure in \textbf{a-f} and that can explain both the energy and the gate dependence of the Kerr data in Fig.\,\ref{fig2}. See text for a detailed explanation.}
	\label{fig3}
\end{figure*}

\subsection{Deviation from the expected type-II band alignment}
\begin{figure*}[tb]
	\includegraphics[width=\linewidth]{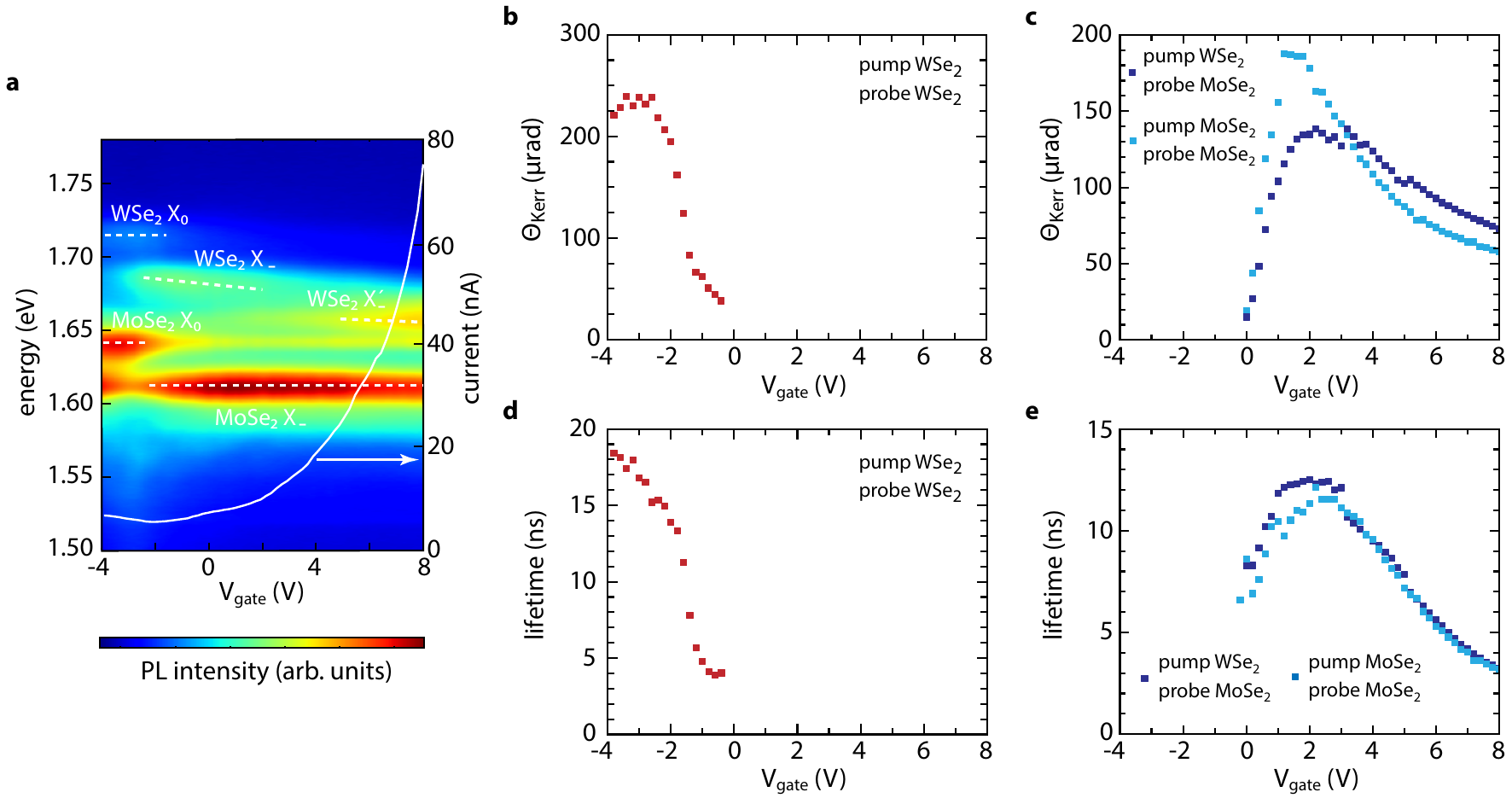}
	\caption{\textbf{Photoluminescence and time-resolved Kerr rotation measurements on a third WSe$_2$/MoSe$_2$ heterobilayer with large twist angle (LTA device~\#3).} \textbf{a} The gate-dependent PL map of LTA device~\#3 that is plotted on a logarithmic color scale is also showing intralayer exciton emission of WSe$_2$, comparable to LTA device \#1 in Fig.\,\ref{fig2}\textbf{a}. Because LTA device \#3 was built with several graphite contacts, electrical transport measurements could be conducted on this device. The white line depicts the current at a bias voltage of \unit[2]{V}. \textbf{b} The bound-exciton related Kerr rotation signal has the highest amplitude at the gate voltage range at which the neutral exciton emission in \textbf{a} indicates the position of the band gap. \textbf{c} The Kerr signal that can be attributed to a valley polarization of free charge carriers appears as soon as the current in \textbf{a} starts to rise at a gate voltage of around $V_\text{gate} = \unit[0]{V}$. \textbf{d,e} Lifetimes of the signals in \textbf{b} and \textbf{c}, respectively. All measurements were conducted at \unit[10]{K}, pump and probe energies are the trion energies of the specified TMD.}
	\label{fig4}
\end{figure*}

To explain this intriguing charge transfer process, we first discuss the PL data in Fig.\,\ref{fig2}a in more detail to determine the band structure of LTA device \#1 and the gate-dependent position of the Fermi level. There is a striking gate-tunability of the WSe$_2$ trion emission, which shows a pronounced red-shift and a transition from the $X_-$ to the $X'_-$ trions, which is normally seen in monolayers of WSe$_2$~\cite{NatNano.8.634,NanoLett.17.740}, but to our knowledge has not yet been reported in such clarity in WSe$_2$/MoSe$_2$-heterobilayers. The exact origin of the $X'_-$-feature is still under debate: Whereas in earlier publications it was attributed to the filling of the energetically higher, spin-split conduction band of WSe$_2$, more recent publications explain this feature with a coupling between plasmons and excitons or the creation of many-body exciton states ~\cite{PhysRevLett.120.066402,PRX.7.041040,PhysRevB.99.085301,PhysRevLett.129.076801}. Whatever its exact origin is, the appearance of this feature is only expected if the conduction band of WSe$_2$ at the K-point gets filled with electrons. However, this is unexpected in case of a type-II WSe$_2$ to MoSe$_2$ band alignment with well-separated conduction band minima of WSe$_2$ and MoSe$_2$ (see Fig.\,\ref{fig1}c)~\cite{PRB.97.165306,Phys.Rev.B.97.205417,ACSNano.12.2498,ACSNano.12.4719,NanoLetters.18.5146}. It was proposed that large displacement fields or high gate-induced charge carrier densities can reduce the energy separation between the conduction band minima of WSe$_2$ and MoSe$_2$ at the K-point, which eventually may even lead to a transition from a type-II to a type-I band alignment~\cite{SolidStateCommunications.266.1115,NatureComm.11.2640,PRB.94.241303}. However, this explanation cannot be applied to LTA device \#1 as the trion emission of both MoSe$_2$ and WSe$_2$ in Fig.\,\ref{fig2}a appear almost simultaneously at gate voltages that are only slightly larger than \unit[0]{V}. Hence, it seems that the conduction band minima of MoSe$_2$ and WSe$_2$ are aligned close to each other, which allows the Fermi level $E_\text{F}$ to almost simultaneously enter both conduction bands (see schematics in Figs.\,\ref{fig3}a and \ref{fig3}d, the blue lines in the figures represent spin-up and red lines represent spin-down bands, slightly opaque rectangles represent the filling of the respective bands with charge carriers). We propose that this unexpected band alignment might be caused by a combination of the large twist angle together with a slight $n$-doping of both the WSe$_2$ and MoSe$_2$ (note that in all PL maps, i.e. Figs.~\ref{fig1}b, \ref{fig1}g, \ref{fig2}a, and \ref{fig4}a, the strongest emission of the neutral exciton is observed at gate voltages $V_\text{gate}<\unit[0]{V}$, especially in Figs.~\ref{fig1}g and \ref{fig4}a it is clear that the negatively charged trion emission prevails over the neutral exciton emission at $V_\text{gate}=\unit[0]{V}$). The donor states may originate either from chalcogenide vacancies or substitutional dopants \cite{AppliedPhysicsLetters.122.160504,NanoResearch.14.1668}. Before stacking the TMD layers into a heterostructure, the respective Fermi level in each TMD layer is therefore not in the middle of the band gap, but somewhere between the donor states and the conduction band minimum. Similar to conventional bulk semiconductors, we expect that the band alignment in the TMD heterostructure will be determined in first order approximation (i.e. ignoring any charge transfer processes) by the alignment of these two Fermi levels. Assuming similar donor binding energies (black solid lines below the conduction bands in Figs.\,\ref{fig3}a and \ref{fig3}d represent donor states) we therefore conclude that the conduction band minima of both TMD layers must be close to each other.

For the given band alignment we expect that for negative gate voltages the Fermi level first enters the valence band of MoSe$_2$, due to the smaller band gap of MoSe$_2$ compared to WSe$_2$ (see Fig.\,\ref{fig3}d). Ignoring any type of band renormalization that might be caused by photo-exited or gate-induced charge carriers~\cite{NatureComm.11.2640,NaturePhotonics.9.466,ACSNano.11.12601}, the valence band maximum of WSe$_2$ in Fig.\,\ref{fig3}d cannot be reached within a technically achievable gate voltage range. This is indeed in accordance to the PL data of Fig.\,\ref{fig2}a that shows the appearance of the positively charged trion ($X_+$) of MoSe$_2$ for $V_\text{gate}<\unit[-1]{V}$, but no clear tuning into the positively charged trion regime of WSe$_2$ (note that we have chosen a logarithmic color scale for all PL maps, the very small emission in the trion energy range of WSe$_2$ can be readily attributed to strain-induced band fluctuations at bubbles in the heterostructure~\cite{AdvancedMaterials.28.9378,LightScienceApplications.9.190}).

Whereas the PL measurements in Fig.\,\ref{fig2}a provide information about the respective band alignment at the K-valleys, we employ ARPES measurements to get further information about the relative position of the $\Gamma$-valley with respect to the K-valleys. Depending on stacking order, interlayer distance, or twist angle, theoretical studies come to different conclusions about whether the valence band maximum lies either at the $\Gamma$- or the K-valley~\cite{PRB.97.165306,Phys.Rev.B.97.205417,ACSNano.12.2498,ACSNano.12.4719,PhysRevResearch.3.043217,Supplement}. Interestingly, even experimental ARPES studies come to different conclusions: Whereas one of the first reported measurements claims that the $\Gamma$-valley lies below the K-valley~\cite{ScienceAdvances.3.e1601832}, a more recent measurement claims the opposite~\cite{Nanotechnology.34.045702}. As we cannot conduct ARPES measurements on the devices with a top hexagonal boron nitride (hBN) layer (see Fig.\,\ref{fig1}a), we fabricated another heterobilayer (LTA device \#2) with a twist angle comparable to LTA device \#1. We used the same crystals and fabrication methods that were used for the other devices and placed a MoSe$_2$/WSe$_2$/graphite-stack directly onto an $n$-doped silicon wafer (i.e.\ with MoSe$_2$ on top, which therefore contributes most to the surface-sensitive ARPES measurements).

The ARPES measurements are shown in Figs.\,\ref{fig3}b and \ref{fig3}e for both s- and p-polarized light with a photon energy of \unit[50]{eV}. The answer to the question in which valley the valence band maximum lies is made more difficult by the fact that the signal strength of an individual valley state depends on the polarisation. Whereas the $\Gamma$-valley can be measured quite well with p-polarized light (Fig.\,\ref{fig3}b), the K-valleys are better visible with s-polarized light (Fig.\,\ref{fig3}e). Only considering linecuts that are normalized and averaged over a small wave-vector range at the K-, K'-, and $\Gamma$-points of the ARPES measurements with p-polarized light therefore may lead to the erroneous conclusion that the $\Gamma$-valley is slightly higher in energy than the K-valleys (see Fig.\,\ref{fig3}c). Instead, considering both the s- and p-polarized measurements and comparing linecuts that show the highest intensity for each individual valley demonstrates that the valence band maximum is rather located at the K-valleys (see Fig.\,\ref{fig3}f). However, $\Gamma$- and K-valleys are energetically close enough to each other that phonon-mediated scattering between these valleys is most likely more relevant than it already is in case of monolayer TMDs~\cite{NanoLetters.19.4083,PRL.119.187402,NanoLetters.18.6135,PhysRevB.101.201405,NatNanotechnol.13.994,NanoLetters.17.4549,NanoLetters.5.2165}.

Based on these ARPES results, we assume that in the other LTA devices the $\Gamma$-valley is also energetically lower than the K-valleys of MoSe$_2$, however energetically close enough such that the $\Gamma$-valley lies in between the K-valleys of WSe$_2$ and MoSe$_2$ (see schematic in Fig.\,\ref{fig3}g). It is also important that our DFT calculations show that the $\Gamma$-valley is spin-degenerate, whereas the Q-valley is showing a small spin-splitting (see Fig.\,\ref{fig1}c and Supplementary Information~\cite{Supplement}), which is consistent to previous studies~\cite{PRB.97.165306,ACSNano.12.2498,ACSNano.12.4719,PhysRevB.97.245427}.

\subsection{Model of the interlayer transfer mechanism}
With this knowledge, we now provide a model that explains the observed dependence of the valley polarization (Kerr rotation amplitude) in Figs.\,\ref{fig2}d to \ref{fig2}f on both the pump/probe energy and the gate. Figs.\,\ref{fig3}g to \ref{fig3}j show the underlying dynamics for the case that the energy of the pump laser is set to WSe$_2$ and the Fermi level enters the valence band of MoSe$_2$, i.e.\ $\unit[-6]{V}<V_\text{gate}<\unit[-1]{V}$ in Fig.\,\ref{fig2}. Under such conditions, the right circularly polarized pump pulse will excite charge carriers from the upmost valence band of WSe$_2$ to its upper spin-split conduction band (see transition from Fig.\,\ref{fig3}g to Fig.\,\ref{fig3}h)~\cite{RevModPhys.90.021001}. It is likely that the predominant energy relaxation path for the photo-excited electrons involves an interlayer transfer from the upper spin-split conduction band of WSe$_2$ to the lower spin-split conduction band of MoSe$_2$, with the same spin-orientation and the same K-valley (see upper black dashed line in Fig.\,\ref{fig3}h). In contrast, the photo-excited holes can experience spin scattering caused by the spin-degenerate $\Gamma$-valley (see transition from Fig.\,\ref{fig3}h to Fig.\,\ref{fig3}i) leading to a population of the K'-valley's valence band with photo-exited holes.

The different interlayer spin/valley-scattering paths and rates of the photo-excited electrons and holes lead to a situation where eventually more photo-exited electrons than photo-excited holes are in the K-valley of MoSe$_2$, whereas the situation is reversed in the K'-valley (see Fig.\,\ref{fig3}i). During exciton recombination (see transition from Fig.\,\ref{fig3}i to Fig.\,\ref{fig3}j), photo-excited electrons also recombine with free holes in the K-valley, pushing the charge carrier density below the Fermi level (continuous dashed black line through Figs.\,\ref{fig3}g to Fig.\,\ref{fig3}j). As there are more photo-excited holes than electrons in the K'-valley, it now shows a surplus of holes after the recombination process. Overall, this mechanism transfers the valley polarization from excitons to free charge carriers by a valley selective interlayer charge transfer. The resulting valley polarization then relaxes back into the equilibrium state with the genuine valley lifetime $\tau_\text{valley}^\text{VB}$ (see Fig.\,\ref{fig3}j).
 
The same interlayer transfer mechanism can be readily applied to the case in which the Fermi level is tuned into the conduction band of MoSe$_2$, resulting in a net valley polarization of free charge carriers in the conduction band. As the Q-valley plays a less significant role in the scattering process than the $\Gamma$-valley, we expect, however, that the respective valley lifetimes in the conduction band are longer than the ones in the valence band, which is indeed observed in our data (see Figs.\,\ref{fig2}h and \ref{fig2}i). Our model can also explain the great similarity between Figs.\,\ref{fig2}e,h on the one hand and Figs.\,\ref{fig2}f,i on the other hand, i.e.\ between setting the pump energy either to WSe$_2$ or MoSe$_2$. Creating electron-hole pairs directly in the K-valley of MoSe$_2$ would lead to the exact same condition that is depicted in Fig.\,\ref{fig3}i, if we assume that the photo-excited holes will be subjected to scattering via the $\Gamma$-valley. Despite the energy separation of the $\Gamma$-valley from the K-valleys of MoSe$_2$, this scattering is very likely to occur during the short timescale of the pump pulse because of photo-induced band gap renormalization effects and the amount of available phonons that are created by the scattering of photo-excited charge carriers into lower-energy states~\cite{NanoLetters.17.644,NaturePhotonics.9.466,ACSNano.11.12601,NaturePhysics.14.355}.

Finally, the long-lived Kerr rotation signal that is present when the Fermi level lies within the band gap of one of the two TMD layers (reddish data points in Figs.\,\ref{fig2}d to \ref{fig2}f) can be explained by the mechanism that was discussed in Ref.~\cite{NanoLetters.20.3147} and can be easily incorporated into our current model: A part of the photo-excited electron-hole pairs scatter into defect-bound exciton states (see Fig.\,\ref{fig3}h), which can have quite long recombination times $\tau_\text{rec}^\text{bound}$ from the nanosecond up to the microsecond range~\cite{PhysRevB.96.121404,PhysRevLett.121.057403,NatureCommunications.12.871} and also may create a valley polarization of resident charge carriers during their recombination process (Fig.\,\ref{fig2}j). As this polarization process is directly connected to the lifetime of the bound excitons, i.e.\ the polarization occurs during the full exciton recombination time, it does not allow the direct probing of the respective valley lifetimes (see rate equation model in the Supplemental Material of Ref.~\cite{NanoLetters.20.3147} for a more detailed discussion).  

We note that we simplified our model in Figs.\,\ref{fig3}g to \ref{fig3}j in the sense that we did only consider the most important scattering and recombination channels. However, taking e.g.\ additional spin-flip scattering processes or non-radiative recombination processes into account does not change the overall result, as long as the photo-excited electrons and holes undergo differently strong valley scattering processes. Here a simplified explanation for this necessary condition: If the photo-excited holes recombine with the photo-excited electrons in the same valley (either the valley in which they were created or in which both were scattered into) the system will end up in its initial state, which means that no net valley polarization of free charge carriers can be induced. Instead, with different scattering rates for electrons and holes, there will be an imbalance of photo-excited charge carriers in the conduction and valence bands of a specific valley and, therefore, their recombination necessitates the involvement of resident (free) charge carriers, which get polarized during the recombination process.

\subsection{Counter-play of two different scattering channels}
Finally, we explore another device with large twist angle (LTA device \#3), which additionally allows for charge transport measurements as it was built with multiple graphite contacts. The gate-dependent PL map of this device (Fig.\,\ref{fig4}a) is quite comparable to the one of LTA device \#1 in Fig.\,\ref{fig2}a, especially with respect to the appearance of the exciton emission of WSe$_2$. When tuning both pump and probe energies to WSe$_2$ we again observe the long-lived Kerr rotation signal (Figs.\,\ref{fig4}b and \ref{fig4}d) that is related to the bound exciton states and that has its largest amplitude at the gate voltage range where the neutral exciton of WSe$_2$ is most pronounced in Fig.\,\ref{fig4}a.  

The white solid line in Fig.\,\ref{fig4}a depicts the source-drain current measured at a bias voltage of \unit[2]{V} and shows that the current starts to rise at a gate voltage of around $V_\text{gate} = \unit[0]{V}$. It is exactly at this gate voltage at which the TRKR signal that can be attributed to the valley polarization of free charge carriers appears when the probe energy is tuned to MoSe$_2$ (see Fig.\,\ref{fig4}c). Its amplitude first increases almost linearly with the gate voltage (the more free charge carriers are induced by the gate, the more can be polarized), before it eventually decreases again for larger gate voltages, i.e.\ larger electron densities, which is in complete accordance to LTA device \#1. The respective valley lifetimes are shown in Fig.\,\ref{fig4}e, whereas Fig.\,\ref{fig4}d shows the lifetime of the bound-exciton driven signal. As it was the case for LTA device \#1 (see Figs.\,\ref{fig2}h and \ref{fig2}i), the valley lifetime of the free charge carriers (Fig.\,\ref{fig4}e) first increases for small carrier densities, then exhibit a maximum at intermediate densities, before it eventually decreases towards higher densities. We attribute this overall trend to an interplay of two different scattering channels that have opposite dependencies with respect to the Fermi level. Pushing the Fermi level further into the conduction band reduces the scattering via mid-gap or tail states. However, at the same time, wave-vector dependent electron-phonon and spin-orbit scattering mechanisms get stronger~\cite{NanoLetters.20.3147,PhysRevB.87.245421,PhysRevB.93.035414,NanoLetters.17.4549}.

We note that when probing at the MoSe$_2$ energy, LTA device \#3 does not exhibit any long-lived, bound-exciton related Kerr rotation signal when the Fermi level is tuned into the band gap. This indicates a low defect density in the MoSe$_2$ layer. At the same time, we observe much longer valley lifetimes of free charge carriers compared to LTA device \#1 (up to \unit[12]{ns} instead of \unit[2]{ns}). Combining these two observations supports the notion that mid-gap or tail states limit the genuine valley lifetime for small charge carrier densities.

\section{Discussion}
In summary, we have demonstrated that the twist angle has a significant impact on the dynamics of valley-polarized free charge carriers in WSe$_2$/MoSe$_2$ heterobilayers. For small twist angles near a crystallographic stacking order (an R-type stacking order for a twist angle of zero, an H-type stacking order for a twist angle of $\unit[60]{^\circ}$), scattering via Q- and $\Gamma$-valleys highly diminishes the valley polarization of free charge carriers, despite the simultaneous occurrence of interlayer excitons with their presumably long recombination and polarization times. For twist angles that lie between these stacking orders, we observe a substantial increase in both the magnitude and the lifetime of the valley polarization, hinting to a significant reduction in the scattering via the Q- and $\Gamma$-valleys. Gate-dependent measurements of these latter devices enable us to disentangle two different Kerr signals, one that is related to defect-bound states within the band gap region, and one that represents the actual valley-polarization of free charge carriers within either valence or conduction band. Interestingly, for these devices we also observe a deviation from the widely assumed type-II band alignment that we contribute to an alignment of the respective conduction bands by donor states. This observation could open the possibility of tailoring the band alignment in these heterobilayer devices, and thus the valley and spin dynamics, by using differently doped TMD layers. Most importantly, the unexpected band structure alignment in these devices enables an interlayer transfer of photo-induced, valley-polarized excitons from the WSe$_2$ layer into a pronounced valley polarization of free charge carriers in the MoSe$_2$ layer. This transfer is most efficient at the onsets of either valence or conduction band, demonstrating the possibility of extracting and utilizing this valley polarization by electrical means.

\section{Methods}
\subsection{Device fabrication}
The WSe$_2$/MoSe$_2$-heterobilayer devices are fabricated from exfoliated flakes with a dry-transfer method, protected by hBN from both sides and contacted via graphite electrodes (see schematic in Fig.\,\ref{fig1}a; detailed information about device fabrication can be found in Refs.~\cite{NanoLetters.20.3147,NanoLetters.22.4949}). For STA device \#1 and LTA devices \#1 and \#3 we furthermore incorporated a graphite gate, whereas for STA device \#2 we use the Si$^{++}$/SiO$_2$ (\unit[285]{nm}) wafer for gating. An exception from this general fabrication procedure is LTA device \#2 that is used for the ARPES measurements and therefore cannot be protected by hBN. For this device we directly placed a MoSe$_2$/WSe$_2$/graphite-stack onto an n-doped silicon wafer.

\subsection{Measurements}
Detailed technical information about the time-resolved Kerr rotation and photoluminesence measurement techniques and the used fitting procedures can be found in the Supporting Information sections of Refs.~\cite{NanoLetters.20.3147,NanoLetters.19.4083,PhysRevB.95.235408}. We note that it is important to account for photo-induced screening effects of the gate electric field~\cite{PhysStatusSolidiRRL.14.2000298} to unveil the true lifetimes and amplitudes of the Kerr signal that can be attributed to free charge carriers. In the Supporting Information of Ref.~\cite{NanoLetters.20.3147} we show how the standard optical measurement techniques can be modified to account for the photo-induced gate screening and we demonstrate that disregarding this procedure can lead to erroneous conclusions drawn from gate-dependent measurements. 

\subsection{Time-resolved Kerr rotation}
Two mode-locked Ti:sapphire lasers are used to independently tune the energies of both pump and probe pulses. An electronic delay between both pulses covers the full laser repetition interval of \unit[12.5]{ns}. The pulse widths are on the order of \unit[3]{ps}. The laser beams are focused onto the device by a \unit[15]{mm} focal length aspheric lens (numerical aperture of 0.66 NA). Typical spot diameters are $\unit[6-8]{\mu m}$ measured as a full width at half maximum (FWHM) value. The laser power was around $\unit[500]{\mu W}$ for both pump and probe beams.

\subsection{Photoluminescence}
A microscope objective lens (numerical aperture of 0.5~NA) is used for the photoluminescence measurement. The resulting spot size is around $\unit[1-2]{\mu m}$ FWHM. A continuous wave laser with an energy of \unit[2.33]{eV} and a power between $\unit[10-50]{\mu W}$ is used for these measurements.

\subsection{Angle-resolved photoemission spectroscopy}
The ARPES measurements have been performed at the NanoESCA beamline of Elettra, the Italian synchrotron radiation facility, using a FOCUS NanoESCA photoemission electron microscope (PEEM) in the k-space mapping mode operation~\cite{J.Electron.Spectrosc.Relat.Phenom.185.330}. The PEEM is operating at a background pressure $p < \unit[5\times 10^{-11}]{mbar}$ and the photoelectron signal is collected from a spot size of about $\unit[5-10]{\mu  m}$. Before the experiment the sample was outgassed in UHV at $T=\unit[180]{^\circ C}$ for \unit[60]{min}. The measurements were conducted with a photon energy of \unit[50]{eV} and an overall energy resolution of \unit[50]{meV} using p- and s-polarized synchrotron radiation, while keeping the sample at \unit[90]{K}.

\section{Data availability}
All data are available from the corresponding author upon reasonable request.

\section{Acknowledgements}
The authors thank Riccardo Reho, Pedro Miguel MC de Melo, Zeila Zanolli, and Matthieu Verstraete for helpful discussions.
This project has received funding from the European Union’s Horizon 2020 research and innovation programme under grant agreement No 881603, by the Deutsche Forschungsgemeinschaft (DFG, German Research Foundation) under Germany's Excellence Strategy - Cluster of Excellence Matter and Light for Quantum Computing (ML4Q) EXC 2004/1 – 390534769, and by the Helmholtz Nanoelectronic Facility (HNF) at the Forschungszentrum J\"ulich~\cite{HNF}.
P.E.F.J. and J.F. acknowledge the financial support of the Deutsche Forschungsgemeinschaft (DFG, German Research Foundation) SFB 1277 (Project-ID 314695032, projects B07 and B11), SPP 2244 (Project No. 443416183), and of the European Union Horizon 2020 Research and Innovation Program under Contract No. 881603 (Graphene Flagship). L.W. acknowledges support by the Alexander von Humboldt foundation. K.W. and T.T. acknowledge support from JSPS KAKENHI (Grant Numbers 19H05790, 20H00354 and 21H05233).

\section{Author contributions}
B.B., C.S., and F.V. conceived and supervised the project.
M.E. performed the TRKR, PL, and electrical measurements with the support of F.V., L.R., and S.D.. 
M.E. and F.V. analyzed the data from the TRKR, PL, and electrical measurements.
P.E.F.J. and J.F. performed the first-principles calculations.
L.W. performed and analyzed the SHG.
F.V., M.E., and B.B. derived the model of the interlayer transfer mechanism.
L.R., S.D., and B.P. designed and fabricated the devices.
B.P., L.P., I.C., V.F., and C.M.S. performed, analyzed, and supervised the ARPES measurements.
K.W. and T.T. synthesized the hBN crystals.
F.V. and B.B. wrote the paper with contributions from all authors.

\section{Competing interests}
The authors declare no competing interests.

\end{document}


\title{Supplementary Information:\\Twist angle dependent interlayer transfer of valley polarization from excitons to free charge carriers in WSe$_2$/MoSe$_2$ heterobilayers}

\author{Frank Volmer}
\affiliation{2nd Institute of Physics and JARA-FIT, RWTH Aachen University, 52074 Aachen, Germany}
\affiliation{AMO GmbH, Advanced Microelectronic Center Aachen (AMICA), 52074 Aachen, Germany}

\author{Manfred Ersfeld}
\affiliation{2nd Institute of Physics and JARA-FIT, RWTH Aachen University, 52074 Aachen, Germany}

\author{Paulo E. Faria~Junior}
\affiliation{Institut für Theoretische Physik, Universität Regensburg, D-93040 Regensburg, Germany}

\author{Lutz Waldecker}
\affiliation{2nd Institute of Physics and JARA-FIT, RWTH Aachen University, 52074 Aachen, Germany}
\affiliation{Department of Applied Physics, Stanford University, 348 Via Pueblo Mall, Stanford, CA 94305, USA}

\author{Bharti Parashar}
\affiliation{Peter Gr\"unberg Institute (PGI-6), Forschungszentrum J\"ulich GmbH, 52428 J\"ulich, Germany}

\author{Lars Rathmann}
\affiliation{2nd Institute of Physics and JARA-FIT, RWTH Aachen University, 52074 Aachen, Germany}

\author{Sudipta Dubey}
\affiliation{2nd Institute of Physics and JARA-FIT, RWTH Aachen University, 52074 Aachen, Germany}

\author{Iulia Cojocariu}
\affiliation{Peter Gr\"unberg Institute (PGI-6), Forschungszentrum J\"ulich GmbH, 52428 J\"ulich, Germany}

\author{Vitaliy Feyer}
\affiliation{Peter Gr\"unberg Institute (PGI-6), Forschungszentrum J\"ulich GmbH, 52428 J\"ulich, Germany}
\affiliation{Fakult\"at f\"ur Physik and Center for Nanointegration Duisburg-Essen (CENIDE), Universität Duisburg-Essen, D-47048 Duisburg, Germany}

\author{Kenji Watanabe}
\affiliation{Research Center for Functional Materials, National Institute for Materials Science, Tsukuba 305-0044, Japan}

\author{Takashi Taniguchi}
\affiliation{International Center for Materials Nanoarchitectonics, National Institute for Materials Science, Tsukuba 305-0044, Japan}

\author{Claus M. Schneider}
\affiliation{Peter Gr\"unberg Institute (PGI-6), Forschungszentrum J\"ulich GmbH, 52428 J\"ulich, Germany}
\affiliation{Fakult\"at f\"ur Physik and Center for Nanointegration Duisburg-Essen (CENIDE), Universität Duisburg-Essen, D-47048 Duisburg, Germany}

\author{Lukasz Plucinski}
\affiliation{Peter Gr\"unberg Institute (PGI-6), Forschungszentrum J\"ulich GmbH, 52428 J\"ulich, Germany}

\author{Christoph Stampfer}
\affiliation{2nd Institute of Physics and JARA-FIT, RWTH Aachen University, 52074 Aachen, Germany}
\affiliation{Peter Gr\"unberg Institute (PGI-9), Forschungszentrum J\"ulich, 52425 J\"ulich, Germany}

\author{Jaroslav Fabian}
\affiliation{Institut für Theoretische Physik, Universität Regensburg, D-93040 Regensburg, Germany}

\author{Bernd Beschoten}
\affiliation{2nd Institute of Physics and JARA-FIT, RWTH Aachen University, 52074 Aachen, Germany}
\affiliation{JARA-FIT Institute for Quantum Information, Forschungszentrum J\"ulich GmbH and RWTH Aachen University, 52074 Aachen, Germany}

\maketitle
\tableofcontents

\section{Second Harmonic Generation}
The twist angles in the heterobilayer devices were determined by measurements of the polarization dependence of the second harmonic (SH) generation intensity. In monolayer samples, this dependence immediately indicates the orientation of the crystallographic axes ~\cite{PhysRevB.87.201401,NanoLetters.13.3329}. In a heterobilayer, the SH field is given by a vector superposition of the individual SH fields of both monolayers, which allows to distinguish natural stacking with almost complete destructive interference from the unnatural stacking with constructive interference~\cite{ACSNano.8.2951}.

\begin{figure}[tb]
	\includegraphics[width=\linewidth]{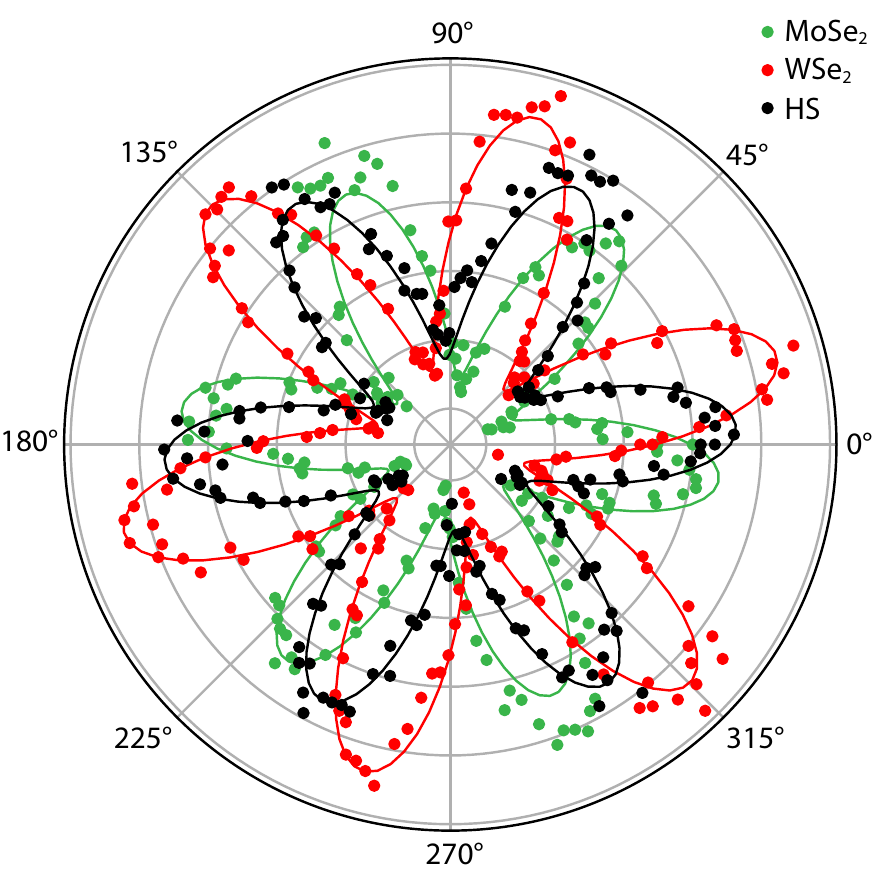}
	\caption{Polar plot of the polarization-dependent second harmonic intensity as a function of the azimuthal angle of the linearly polarized incident laser for LTA device~\#1 of the main manuscript.}
	\label{S1}
\end{figure}

We measured the SH intensity on parts of isolated monolayers of both materials as well as on the bilayer part. Fig.~\ref{S1} shows a representative measurement of the intensities of the SH emission as a function of the azimuthal angle of the linearly polarized incident laser in case of LTA device~\#1 of the main manuscript. The data of the monolayers was fitted with the function $I_{\Theta} = I_0 \cdot \sin^2(3\Theta)$. The difference of the numerically optimized angles $\Theta$ between the different areas is the relative twist angle given in the main manuscript. In the case of small angles between the crystallographic axes (STA devices~\#1 and \#2), the SH intensity of the bilayer was compared to the intensity of both monolayers, which identifies the twist angle of STA device~\#1 as $7^\circ$ and of STA device~\#2 as $56^\circ$, which is a small twist angle of $60^\circ-56^\circ=4^\circ$ to an H-type stacking order.

Due to quite different sizes of the two TMD flakes in LTA device \#3 and the use of the monolayer parts for electrical contacts, we were not able to obtain the necessary SHG measurements on the respective monolayer areas in this device. Therefore, we cannot give a twist angle for this device. However, we note that we intentionally misaligned the edges of the TMD flakes during the stacking process to achieve a large twist angle. The PL and TRKR measurements indicate that the large twist angle was indeed achieved.

The laser used in the experiments had a pulse duration of approximately \unit[200]{fs}, a repetition rate of \unit[100]{MHz} and a central wavelength of \unit[1030]{nm}. The average power used was between \unit[20]{mW} and \unit[40]{mW}. 

\begin{figure*}[tb]
	\includegraphics[width=\linewidth]{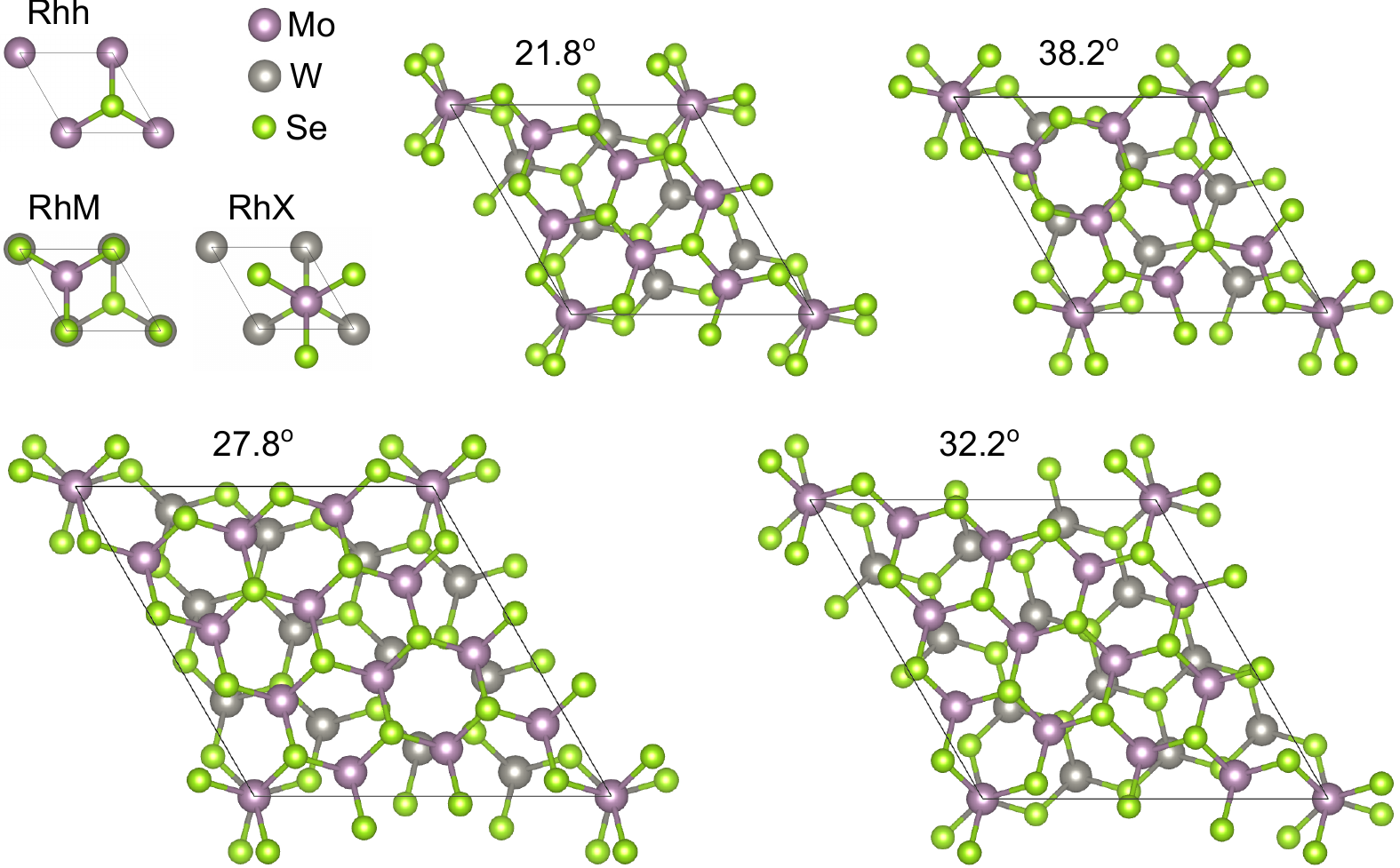}
	\caption{Supercells for R-stacking and large twist angles investigated within first principles.}
	\label{S2}
\end{figure*}

\begin{figure*}[tb]
	\includegraphics[width=\linewidth]{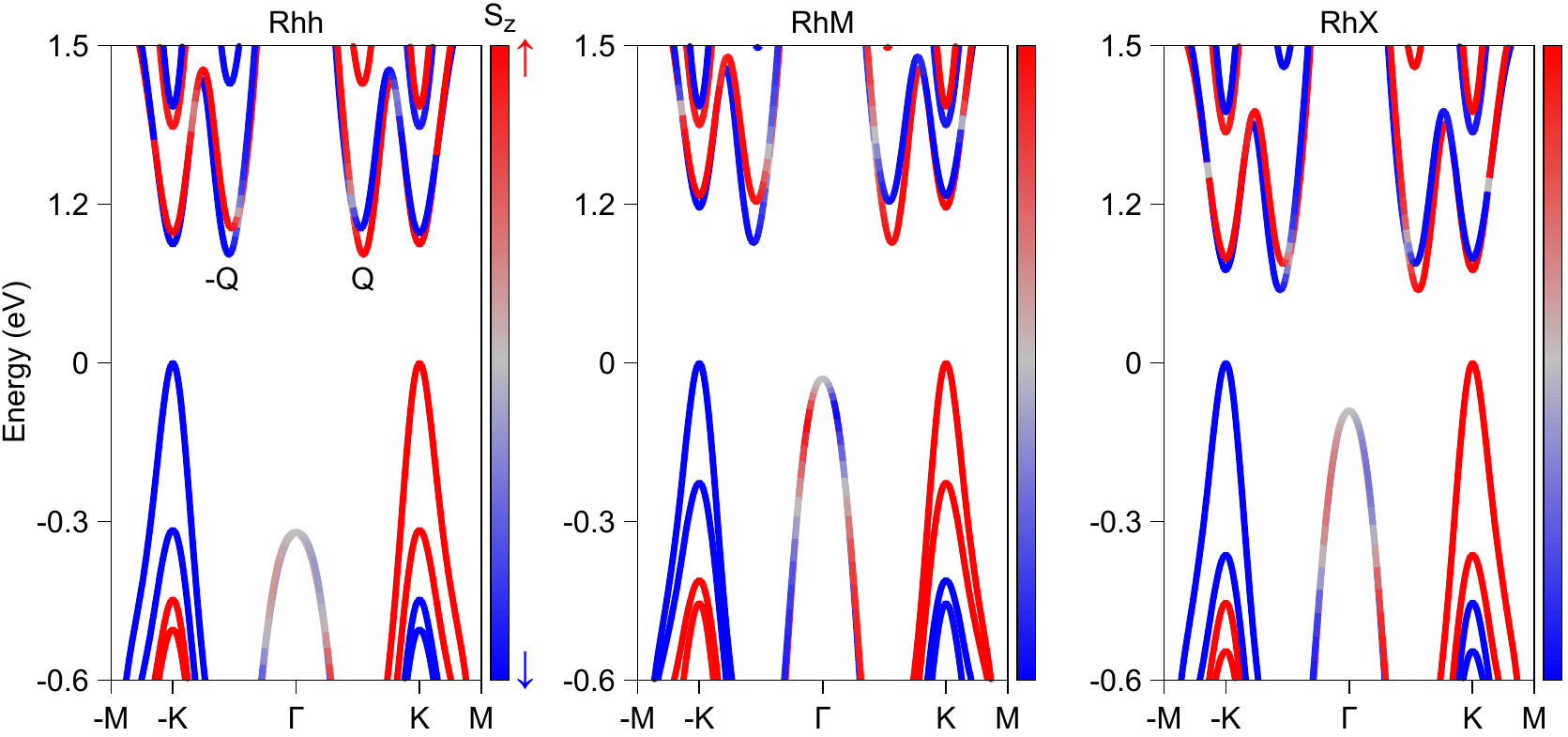}
	\caption{Spin-resolved band structures for the three different R-type stackings at $\unit[0]{^\circ}$ twist-angle.}
	\label{S3}
\end{figure*}

\begin{figure*}[tb]
	\includegraphics[width=\linewidth]{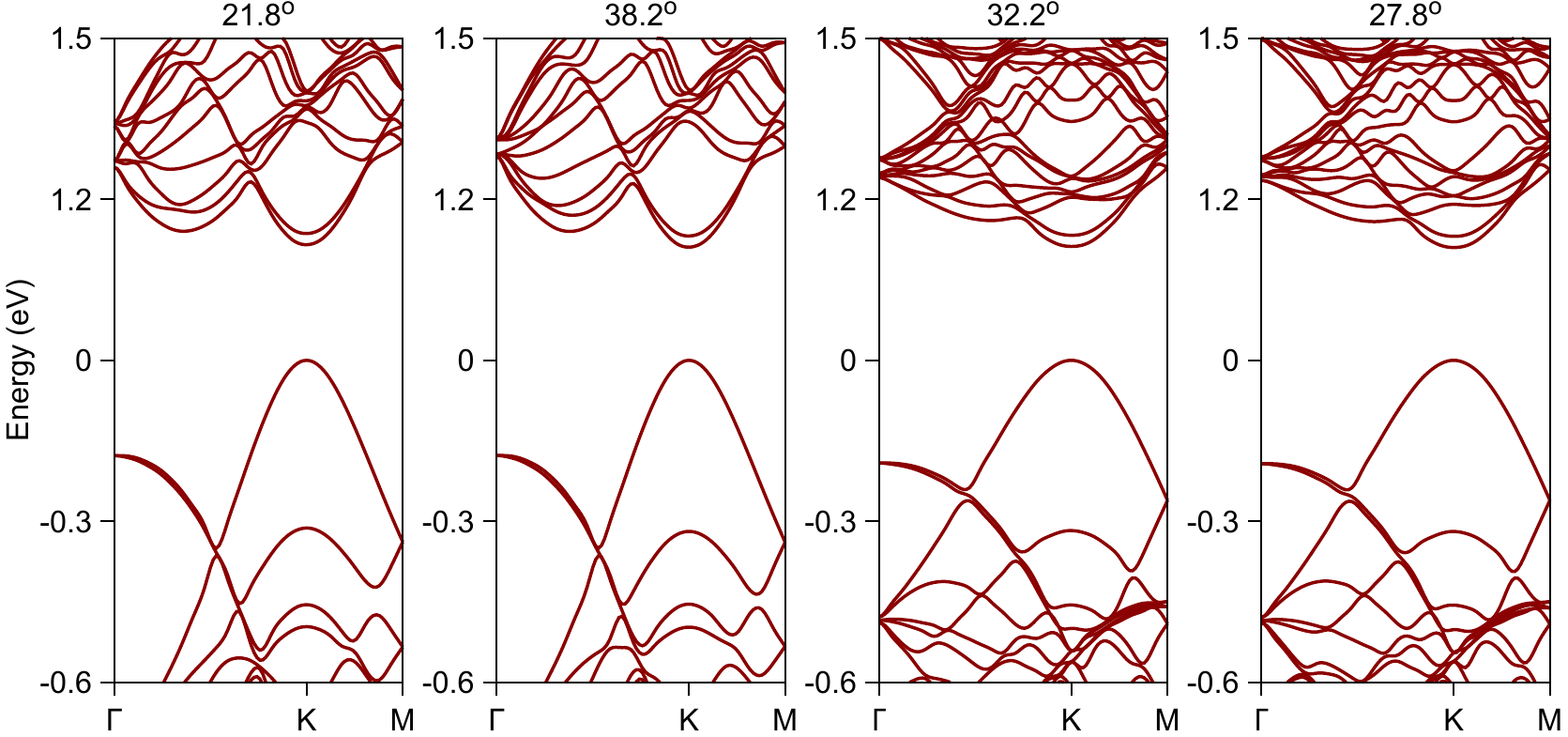}
	\caption{Calculated band structures for large twist angles.}
	\label{S4}
\end{figure*}

\begin{figure*}[tb]
	\includegraphics[width=\linewidth]{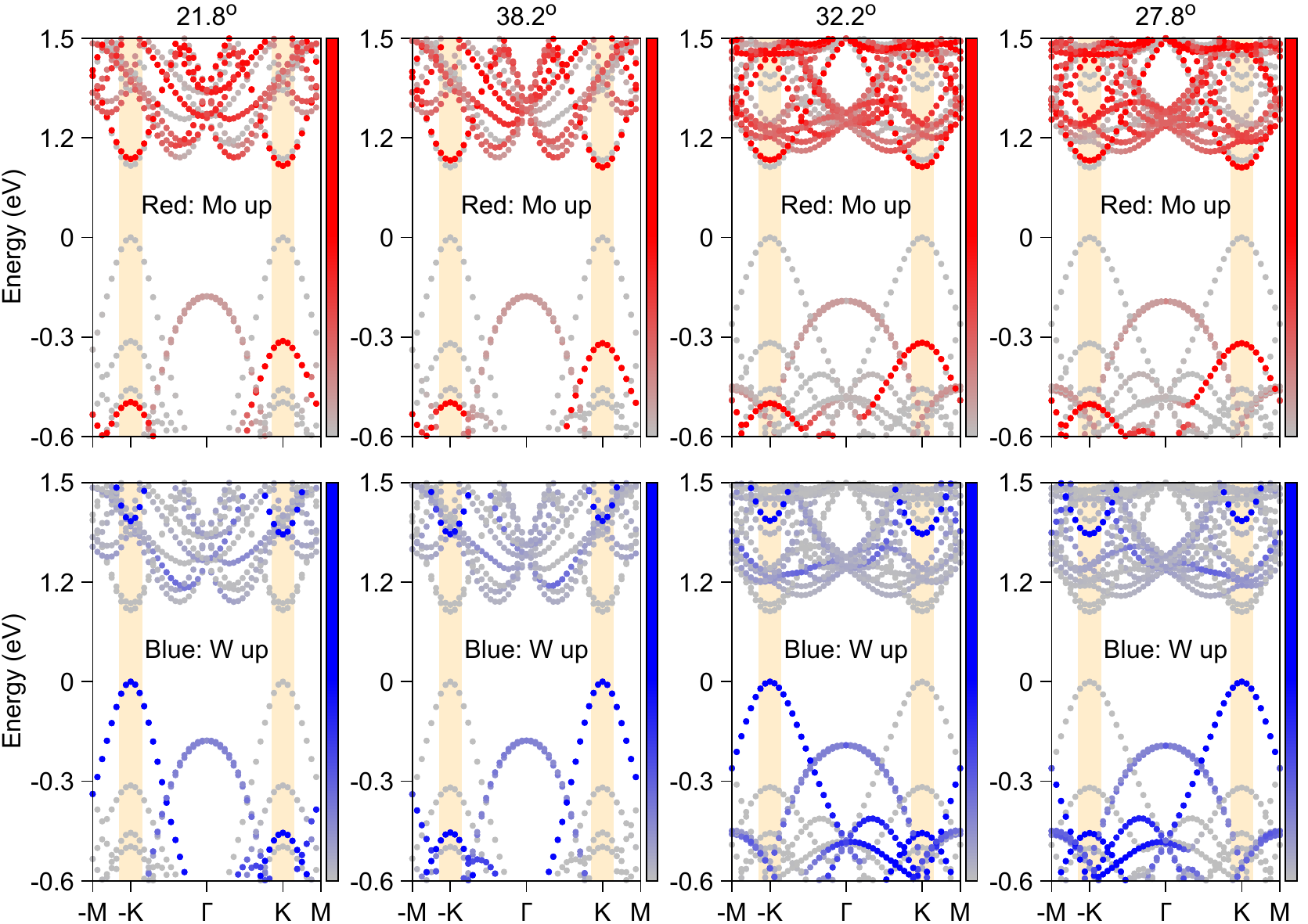}
	\caption{Band structures for large twist angles with highlighted contributions from the spin-up states of molybdenum and tungsten atoms.}
	\label{S5}
\end{figure*}

\section{Sensitivity of Kerr signal to free charge carriers}
\label{SensitivityKerrFreeChargeCarriers}
As we have demonstrated in Ref.~\cite{NanoLetters.20.3147}, time-resolved Kerr rotation is not only sensitive to valley-polarized excitons but can also detect a valley polarization of free charge carriers, if the energy of the probe pulse is tuned to the trion (charged exciton) energy of the corresponding TMD. This is due to the fact that both the helicity of the incoming light and its energy determines what type of trion (singlet or triplet) is formed and in which valley the trion is created ~\cite{RevModPhys.90.021001,NatureComm.7.12715,PhysRevB.95.235408}. As the formation of trions necessitates the presence of additional charge carriers~\cite{RevModPhys.90.021001}, the free charge carrier density inside the valley directly impacts the probability with which photons with matching helicity and energy interact with the corresponding valley. The sensitivity of the linearly polarized probe pulse on a valley polarization can therefore be explained by the fact that the probe pulse is a superposition of left and right circularly polarized light, which have different probabilities to interact with the TMD in the presence of a valley polarization of free charge carriers. Of course, this is a simplified explanation as discussed in the next section.

\section{No sensitivity of Kerr signal to interlayer excitons}
\label{NoSensitivityKerrInterlayerExcitons}
Our observations in the main manuscript raise the question of why TRKR does not seem to be able to detect a polarization of interlayer excitons. The main problem in this respect is that it is everything but trivial to derive a theoretical expression for the expected Kerr signal. This is due to the fact that the devices in our study are complex multilayer structures that consist of magneto‐optic materials (the TMDs) and dielectric materials (hBN and SiO$_2$) placed on a highly reflective substrate (Si). This greatly complicates the derivation of an expression for the Kerr signal, as discussed in Refs.~\cite{JApplPhy.38.1652,JMagnMat.89.107,JApplPhys.84.541}. To our knowledge, there is no theoretical derivation of the Kerr signal for TMD heterobilayers to date.

In the energy range of the interlayer exciton such a derivation might be especially difficult, as it was shown that interlayer excitons at different moir\'e sites can exhibit opposite optical selection rules \cite{NaturePhysics.15.1140,PhysRevB.97.035306,PhysRevLett.115.187002,2DMaterials.5.035021}. Therefore, it is not clear if an interlayer exciton polarization can actually rotate the linear polarization of the probe pulse, i.e. creating a Kerr rotation.

\section{Band structure calculations}
We performed the first principles calculations of the R-stacked (0° twist) and large twist angles (21.8°, 38.2°, 32.2° and 27.8°) MoSe$_2$/WSe$_2$ heterostructures based on the density functional theory using the all-electron full-potential linearized augmented plane-wave method implemented within the Wien2k package~\cite{wien2k}. We used the Perdew–Burke–Ernzerhof~\cite{Perdew1996PRL} exchange-correlation functional with van der Waals interactions accounted via the D3 correction~\cite{Grimme2010JCP}. The wavefunction expansion into atomic spheres takes into account orbital quantum numbers up to 10 and the plane-wave cut-off multiplied with the smallest atomic radii is set to 7. Spin–orbit coupling was included fully relativistically for core electrons, while valence electrons were treated within a second-variational procedure~\cite{Singh2006} with the scalar-relativistic wavefunctions calculated in an energy window up to 2 Ry. Self-consistency was achieved using a two-dimensional Monkhorst–Pack k-grid with 12x12 (6x6) points for the R-stacked (twisted) structure and the convergence criteria of $10^{-5}$ e for the charge and $10^{-5}$ Ry for the energy was used. We built commensurate heterostructures with a common in-plane lattice parameter of 3.2855 $\AA$ for MoSe$_2$ and WSe$_2$ layers (as an average value of their individual lattice parameters~\cite{Kormanyos2015TDM}). This average lattice parameter leads to an effective strain of $\sim 0.1 \%$ to the individual layers, which is sufficiently small to affect the energy- and spin- dependent properties~\cite{Zollner2019PRB,FariaJunior2022NJP}. A vacuum region of 16 $\textrm{\AA}$ was assumed in all cases. We  minimized the energy to find the equilibrium interlayer distances, obtaining 3.724 $\textrm{\AA}$ for the Rhh stacking, 3.074 $\textrm{\AA}$ for the RhM stacking, 3.084 $\textrm{\AA}$ for the RhX stacking, 3.337 $\textrm{\AA}$ for the 21.8° and 38.2° twisted structures and 3.365 $\textrm{\AA}$ for the 32.2° and 27.8° twisted structures. In Fig.~\ref{S2} we present the top view of each crystal structure considered in our study, namely, the Rhh, RhM and RhX stackings (0° twist angle) and the large twist angle cases 21.8°, 38.2°, 32.2° and 27.8°.

\subsection{Small twist angles}
Fig.~\ref{S3} shows the calculated band structures for the three different R-type stackings at $\unit[0]{^\circ}$ twist-angle (see Fig.~\ref{S2} for a description of these stacking orders). Depending on the type of stacking, the positions of the $\Gamma$- and Q-valleys can shift quite significantly with respect to the K-valleys. Furthermore, together with external parameters like interlayer distance~\cite{Kunstmann2018NatPhys} (which might be impacted by fabrication-induced contamination between the TMD layers) or strain~\cite{Zollner2019PRB}, the difference in stacking order might be the reason that STA device~\#2 shows a less pronounced hybridization than STA device~\#1 despite its smaller twist-angle between the crystallographic axes. This is supported by a recent publication by some co-authors, in which the high-symmetry stackings (R- and H-type) are investigated under electric fields and varying interlayer distances \cite{Nanomaterials.13.1187}. In Figs.~1 and 2 of Ref.~\cite{Nanomaterials.13.1187}, R- and H-type stackings do show different energy alignments between K-, $\Gamma$-, and Q-bands, as well as different dependencies with respect to changes in the electric field and interlayer distance (see Figs.~3 and 7 of Ref.~\cite{Nanomaterials.13.1187}).

The color-code in Fig.~\ref{S3} depicts the spin-texture in the out-of-plane direction. Whereas the conduction band minimum at the Q-valley is showing a small spin-splitting, the valence band maximum at the $\Gamma$-valley is spin-degenerated, therefore likely enhancing spin-flip scattering in the valence band as described in the main manuscript.

\begin{table*}[]
\caption{Dipole matrix elements,  $\left|p_{cv}^{\epsilon}\right|=\left|\left\langle v,K\left|\hat{\epsilon}\cdot\vec{p}\right|c,K\right\rangle \right|$ (in $\textrm{eV}.\textrm{\AA}$), and their respective polarization, $\epsilon= \left\{ +,-,z \right\}$, calculated at the K-point of the studied systems.}
\begin{tabular}{c|cc|cc|cc}
\hline
\hline
 & \multicolumn{2}{c|}{intralayer from VB+(Mo)} & \multicolumn{2}{c|}{intralayer from VB+(W)} & \multicolumn{2}{c}{interlayer from VB+(W)}\tabularnewline
 \hline
 & to CB-(Mo) & to CB+(Mo) & to CB-(W) & to CB+(W) & to CB-(Mo) & to CB+(Mo)\tabularnewline
\hline
Rhh   &  4.62955, +  &  0.11176, z  &  0.36211, z  &  5.85025, +  &  0.32198, +  &  0.05107, z \tabularnewline
RhM   &  4.56963, +  &  0.15023, z  &  0.23852, z  &  5.85374, +  &  0.36175, z  &  0.01253, - \tabularnewline
RhX   &  4.64344, +  &  0.12416, z  &  0.41587, z  &  5.83282, +  &  0.28064, -  &  0.15099, + \tabularnewline
21.8°  &  4.63933, +  &  0.12007, z  &  0.34812, z  &  5.79177, -  &  0.00146, z  &  0.01181, - \tabularnewline
38.2°  &  4.65387, +  &  0.09271, z  &  0.30346, z  &  5.81078, +  &  0.02484, +  &  0.00277, z \tabularnewline
32.2°  &  4.65803, +  &  0.09021, z  &  0.29798, z  &  5.84845, -  &  0.00004, z  &  0.00030, - \tabularnewline
27.8°  &  4.65539, +  &  0.11436, z  &  0.25132, z  &  5.84818, +  &  0.00802, +  &  0.00028, z \tabularnewline
\hline
\hline
\end{tabular}
\label{table-S1}
\end{table*}

\subsection{Large twist angles}
Fig.~\ref{S4} depicts the calculated band structures for the several large twist angles analyzed. We note that all of our calculations resulted in a type II band alignment, consistent with previous theoretical studies, but at odds with our experimental data. So far, theoretical studies only show a transition from a type II to a type I band alignment in case of large displacement fields~\cite{SolidStateCommunications.266.1115,PRB.94.241303}. However, the necessary strength of the displacement field lies far beyond our applicable gate voltage range. Furthermore, in our devices we do not have the combination of both a bottom and a top gate, which is necessary to create a displacement field without simultaneously inducing charge carriers into the heterostructure~\cite{AdvancedElectronicMaterials.2200510}.

On the other hand, there is one experimental study that argues to observe a transition from a type II to a type I band alignment in case of large charge carrier densities~\cite{NatureComm.11.2640}. In this respect, we note that to our knowledge all previous band structure calculations of TMD heterostructures (including our own calculations) are done under the assumption of charge neutrality. This limitation is due to an increase in complexity of the models and especially a significant increase in required computational power if doping is also taken into account. Nonetheless, there has been theoretical efforts along this line but still restricted to monolayers~\cite{Steinhoff2014NL,Erben2018PRB}. As most of our devices show a slight n-doping, we therefore propose that the unexpected band alignment, which is derived from our experimental data, might be caused by a combination of the large twist angle together with the presence of donor states in both TMD layers. It has to be the scope of further studies to investigate this idea further.

Our calculations can also shed some light on the observation that most WSe$_2$/MoSe$_2$-heterobilayers (especially the ones with small twist angle) show a stronger quenching of the WSe$_2$ intralayer exciton compared to the MoSe$_2$ intralayer exciton, as mentioned in the main manuscript. Table~\ref{table-S1} contains calculated transition rates for neutral excitons at the K-point for the different R-stackings and large twist angles analyzed. Whereas the transition rate for the interlayer exciton significantly drops towards larger twist angles (explaining the vanishing of the interlayer exciton in PL measurements of devices with large twist-angle), the intralayer transitions are barely effected by the twist angle and are similar to the ones in the respective monolayer case. Most importantly, the two bright exciton transitions of WSe$_2$ [VB+(W)$\rightarrow$CB+(W)] and MoSe$_2$ [VB+(Mo)$\rightarrow$CB-(Mo)] have very comparable transition rates. 

To understand the often missing WSe$_2$ exciton emission in heterobilayers, we now only have to take into consideration that the intensity of a PL signal depends on both the transition rate and the occupation number in the respective states. As the transition rates are very comparable to each other, our calculations therefore suggest that photo-excited charge carriers must scatter quite fast away from either the valence band maximum or the upper, spin-split conduction band minimum of WSe$_2$. We assume that the upper, spin-split conduction band minimum is actually responsible for this effect. To support our assumption, we present in Fig.~\ref{S5} the band structures of the large twist angle cases highlight the contributions of the spin-up states of each layer, denoted by the molybdenum and tungsten atoms. Spin-down states are just the time-reversal partners of spin-up states are are not show for simplicity. 

We observe a magnitude of zone-folded bands near the conduction band minima of WSe$_2$. Because of these bands, we assume that photo-excited electrons quite efficiently scatter away from the conduction band K-valleys of WSe$_2$ and therefore are not available for the optical recombination process. On the other hand, at the K-valley both the valence and conduction bands of MoSe$_2$ are quite well-separated from other bands. Therefore, scattering might take longer, which in turn increases the chance of optical recombination, which makes the corresponding PL signal to a quite prominent one in our measurements.

Whereas these zone-folded bands can explain the significantly quenched PL emission from WSe$_2$ in heterostructures that exhibit the widely assumed type II band alignment, the quite strong PL emission from WSe$_2$ in both LTA devices~\#1 and \#3 demonstrate that these devices do not possess these pronounced zone-folded bands near the conduction band minima of WSe$_2$. It must be the scope of further studies, if an alignment of the conduction band minima of WSe$_2$ and MoSe$_2$ caused by donor states (as derived in the main manuscript) has this effect.

\section{Possible impact of a moir\'{e} potential}
The sub-peaks seen in the interlayer exciton emission in Fig.~1d of the main manuscript were also previously attributed to the presence of a moir\'{e} potential \cite{NatureNanotechnology.17.227,Nature.567.66,NatureCommunications.12.1656}. Therefore, the question arises if a moir\'{e} superlattice and its associated  potential may play a role in the observed valley and spin dynamics discussed in the main manuscript (we assume that the twist angles even for the STA devices are too large for atomic reconstruction \cite{ACSNano.14.4550,NatureNanotechnology.na.na}). First, we note that a moir\'{e} potential might only be relevant for the STA devices, as the large twist angles in the LTA devices result in moir\'{e} periods that are most likely too small to be able to form a moir\'{e} potential that can trap charges \cite{NatureNanotechnology.17.227}. Instead, for small twist angles it was shown that the resulting moir\'{e} potentials can lead to a spatial trapping of charge carriers \cite{NatureMaterials.19.630,Nature.567.66,npj2DMaterialsandApplications.5.67}. However, the associated moir\'{e} excitons can exhibit lifetimes of up to hundreds of nanoseconds \cite{NatureNanotechnology.17.227,Nature.567.66,ACSNano.16.16862,npj2DMaterialsandApplications.5.67,PhysRevLett.126.047401}, which is in stark contrast to the observed short lifetimes observed in the Kerr rotation measurements on the one device that is showing the possible moir\'{e} features in its interlayer exciton emission (compare Figs.~1d and 1e in the main manuscript). Therefore, we conclude that our TRKR measurements are in all likelihood unable to directly detect these excitons, which is in accordance to sections~\ref{SensitivityKerrFreeChargeCarriers} and ~\ref{NoSensitivityKerrInterlayerExcitons} of this Supplementary Information and the conclusion of the main manuscript that we primarily probe the valley polarization of free charge carriers.

Furthermore, the vast majority of studies examining the effects of a moir\'{e} potential in TMD-based heterobilayers have conducted their experiments on samples without a gate. Therefore, very little is known about the effects of a moir\'{e} potential when the Fermi level is tuned in either the conduction or valence band. However, it is at the band edges where we observe our most important finding, namely the interlayer transfer of a valley polarization from excitons to valley-polarized free charge carriers. Additionally, we note that all experiments on moir\'{e} excitons are performed with very low laser powers, usually in the lower $\unit[]{\mu W}$ range for cw experiments and even in the $\unit[]{nW}$ range for time-resolved PL measurements with pulse widths of at least tens of picoseconds. Instead, TRKR measurements are usually performed with pulse widths in the lower ps range (in our case about \unit[3]{ps}) or even in the higher fs range with laser powers of up to hundreds of $\unit[]{\mu W}$. Since laser spot sizes are quite comparable between most publications (around $\unit[1-10]{\mu m}$ FWHM), the total amount of photoexcited carriers is much higher in TRKR measurements compared to studies investigating moir\'{e} effects. However, a high density of photo-excited or gated-induced charge carriers can cause band renormalization effects~\cite{NatureComm.11.2640,NaturePhotonics.9.466,ACSNano.11.12601}. Whether the energy scale of the moir\'{e} potential is actually relevant under such conditions must be the scope of further studies.

\section{Additional data}
\begin{figure}[tbh]
	\includegraphics[width=0.9\linewidth]{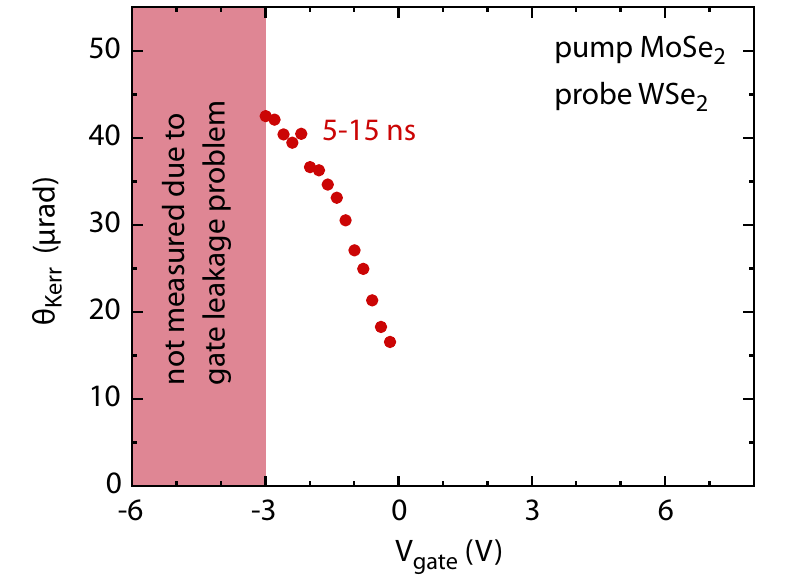}
	\caption{Gate-dependent TRKR amplitude of the band-gap related signal in WSe$_2$ when the pump energy is set to the trion energy of MoSe$_2$ and the probe energy is set to the trion energy of WSe$_2$. The measurement was conducted at \unit[10]{K}.}
	\label{S6}
\end{figure}

Fig.~\ref{S6} shows the Kerr rotation amplitude of the band-gap related signal in LTA device~\#1 when pumping at the trion energy of MoSe$_2$ and probing at the trion energy of WSe$_2$. Thus, it is the last of the four possible combinations of pump and probe energies, while the others are shown in Fig.~2 of the main manuscript. It was also the last energy combination that we measured, as the leakage current over the hBN gate dielectric dramatically increased during this measurement. To avoid a breakthrough of the gate dielectric that would made any further gate-dependent measurements impossible, we stopped the measurement at $V_\text{gate} = \unit[-3]{V}$. However, up to this gate voltage the gate-dependence of the Kerr amplitude is in very good agreement to the one that is shown in Fig.~2d of the main manuscript, where the pump and probe energies are both set to the trion energy of WSe$_2$. This is in complete accordance to our model in Fig.~3 of the main manuscript as the energies of the bound exciton states in WSe$_2$ (dashed line in the band gap of Fig.~3h) are energetically lower than the bright intralayer exciton energies in MoSe$_2$ that are primarily excited by the pump pulse. Through scattering events the pump pulse can therefore occupy the bound excitons states in the band gap of WSe$_2$, creating the long-lived Kerr signal when the probe is set to the trion energy of WSe$_2$.

We note that in our study we show only the amplitudes and lifetimes of the Kerr signals that we attribute to either the long-lived bound exciton states or the valley polarization of free charge carriers. In several of our measurements we also observe a very short-lived signal with a lifetime of up to a few tens of picoseconds. Sometimes this signal is so short-lived that we can only observe it in the first two or three data points after zero delay, which is not enough for a reliable fit. We attribute this signal to a polarization of bright intralayer excitons before they scatter and recombine. As such a polarization of bright intralayer excitons is not the scope of our study, we have excluded the amplitudes and lifetimes of this signal (in case where we were actually able to fit them) from the figures in this study. The fact that no data points for gate voltage above $V_\text{gate} = \unit[0]{V}$ are presented in Fig.~2d of the main manuscript or Fig.~\ref{S6} is due to the fact that no signal besides this very short-lived bright exciton signal is present in these gate voltage ranges.

\section{Spatial positioning of the laser beams}
For the TRKR measurements, we use a single lens to focus the two spatially separated pump and probe beams on the same spot of the device (see Fig.~10 in the Supplemental Material of Ref.~\cite{PhysRevB.95.235408}). The advantage of this approach is that the two beams can be separated quite easily after they have been reflected from the sample (see aforementioned figure in Ref.~\cite{PhysRevB.95.235408}). The disadvantage is that the minimum spot size is limited compared to the use of a microscope objective (the latter does not have a large enough entrance pupil diameter for two spatially separated laser beams). Using a single lens therefore limits the spot diameters to $\unit[6-8]{\mu m}$ measured as the full width at half maximum (FWHM).

The TRKR data therefore probes an area of tens of $\mu m^2$. As typical dimensions of TMD-based heterobilayer devices are at most a few tens of micrometers in length, we therefore search for the one spot in our devices that is furthest away from edges, folds and bubbles in the heterostructure to diminish contributions from these features to the largest possible extent. All measurements (TRKR and PL) are subsequently conducted on this single spot.

\section{Band alignment of LTA devices}
As explained in the main manuscript, our PL measurements on the LTA devices reveal that the conduction band minima of MoSe$_2$ and WSe$_2$ seem to be aligned close to each other, most likely due to an n-doping of the constituent TMD layers. In the schematics of Fig.~3 of the main manuscript we draw the conduction band minimum of MoSe$_2$ slightly below the one of WSe$_2$. Our reason for this is that in the electron regime of LTA device \#1 and \#3 ($V_\text{gate} > \unit[1]{V}$ in Fig.~3 and $V_\text{gate} > \unit[0]{V}$ in Fig.~4 of the main manuscript), we do not observe a Kerr signal that we can attribute to a valley polarization of free charge carriers in WSe$_2$ if we probe at the trion energy of WSe$_2$ (see absent blue data points in Figs.~2d and 4b of the main manuscript). We therefore conclude that the conduction band minimum of MoSe$_2$ has to be slightly lower so that the polarized free charge carriers can occupy only these states, resulting in the Kerr rotation signal that we can attribute to the valley polarization of free charge carriers when probing at the trion energy of  MoSe$_2$ (see blue data points in Figs.~2e and 4c in the main manuscript).